\documentstyle[12pt,aaspp4,psfig]{article}
\begin{document}

\title{The Distribution of Pressures in a Supernova-Driven
Interstellar Medium}

\author{Mordecai-Mark Mac Low\altaffilmark{1}}
\author{Dinshaw Balsara\altaffilmark{2,3}}
\author{Miguel A. Avillez\altaffilmark{1}}
\author{Jongsoo Kim\altaffilmark{2,3,4}}
\altaffiltext{1}{Department of Astrophysics, American Museum of
Natural History, Central Park West at 79th Street, New York, NY,
10024-5192, USA; E-mail: mordecai@amnh.org, mavillez@amnh.org}
\altaffiltext{2}{National Center for Supercomputer Applications,
University of Illinois at Urbana-Champaign, Urbana, IL, 61801, USA;
E-mail: dbalsara@ncsa.uiuc.edu, jskim@ncsa.uiuc.edu} 
\altaffiltext{3}{Department of Physics, Notre Dame University, 225
Nieuwland Science Hall, Notre Dame, IN 46556-5670, USA}
\altaffiltext{4}{Korea Astronomy Observatory, 61-1, Hwaam-Dong,
Yusong-Ku, Taejon 305-348, Korea}


\begin{abstract}
Observations have suggested substantial departures from pressure
equilibrium in the interstellar medium (ISM) in the plane of the
Galaxy, even on scales under 50~pc.  Nevertheless, multi-phase models
of the ISM assume at least locally isobaric gas.  The pressure then
determines the density reached by gas cooling to stable thermal
equilibrium.  We use two different sets of numerical models of the ISM
to examine the consequences of supernova driving for interstellar
pressures.  The first set of models is hydrodynamical, and uses
adaptive mesh refinement to allow computation of a $1\times 1\times
20$~kpc section of a stratified galactic disk.  The second set of
models is magnetohydrodynamical, using an independent code framework,
and examines a (200~pc)$^3$ periodic domain threaded by magnetic
fields.  Both of these models show broad pressure distributions with
roughly log-normal functional forms produced by both shocks and
rarefaction waves, rather than the power-law distributions predicted
by previous work, with rather sharp thermal pressure gradients.  The
width of the distribution of the logs of pressure in gas with $\log T
< 3.9$ is proportional to the rms Mach number in that gas, while the
distribution in hotter gas is broader, but not so broad as would be
predicted by the Mach numbers in that gas.  Individual parcels of gas
reach widely varying points on the thermal equilibrium curve: no
unique set of phases is found, but rather a dynamically-determined
continuum of densities and temperatures.  Furthermore, a substantial
fraction of the gas remains entirely out of thermal equilibrium.  Our
results appear consistent with observations of interstellar pressures,
and suggest that the pressures observed in molecular clouds may be due
to ram pressure rather than gravitational confinement.
\end{abstract}


\clearpage

\section{Introduction}

Theoretical models of the interstellar medium (ISM) have generally
followed Spitzer (1956) in assuming that the interstellar gas is in
pressure equilibrium.  Field, Goldsmith \& Habing (1969; hereafter
FGH) demonstrated that the form of the interstellar cooling and
heating curves for temperatures below about $10^4$~K allowed a range
of pressures in which two isobaric phases could exist in stable
thermal equilibrium.  Although the details of this model turned out to
be incorrect, due to their assumption of a cosmic ray flux much higher
than subsequently observed, Wolfire et al.\ (1995) demonstrated that
photoelectric heating of polycyclic aromatic hydrocarbons or other
very small dust grains could recover a two-phase model for the neutral
ISM.  McKee \& Ostriker (1977; hereafter MO) used a similar framework
for the cold ISM, but incorporated the suggestion by Cox \& Smith
(1974) that supernovae (SNe) would create large regions of hot gas
with $T\sim 10^6$~K, to produce a model of a three-phase medium.  They
were the first to relax the assumption of global pressure equilibrium,
noting that SN remnants (SNRs) would have varying pressures.  They
assumed only local pressure equilibrium at the surfaces of clouds in
order to determine conditions there.

Measurements of the ISM pressure were performed using {\em Copernicus}
observations of ultraviolet (UV) absorption lines from excited states of
C~{\sc i} by Jenkins \& Shaya (1979) and Jenkins, Jura \& Loewenstein
(1983).  They found greater than an order of magnitude variation in
pressures in the cold gas traced by the C~{\sc i}, with most of the
gas having pressures $P/k < 10^4$~K~cm$^{-3}$, but a small fraction
reaching pressures $P/k > 10^5$~K~cm$^{-3}$.  These findings have
recently been confirmed and extended using {\em Space Telescope
Imaging Spectrograph} (STIS) observations by Jenkins \& Tripp (2001).
Bowyer et al.\ (1995) compared the pressure derived from the
observation with the {\em Extreme UV Explorer} of a shadow cast by a cloud
of neutral hydrogen, with the pressure of the warm cloud surrounding
the Sun.  They found the local pressure of 730~K~cm$^{-3}$ to be a
factor of 25 lower than the average pressure measured along the 40 pc
line of sight to the interstellar cloud of $1.9 \times
10^4$~K~cm$^{-3}$.  McKee (1996) argued that these estimates were too
extreme, but nevertheless concluded that the observations showed a
pressure variation of at least a factor of five.

In this paper we show that dynamical models of a SN-driven
interstellar medium do not produce an isobaric medium controlled by
thermal instability, but rather a medium with a broad pressure
distribution controlled by the dynamics of turbulence, with
rarefaction waves being as important as shocks in setting local
pressures.  Although the cooling curve may determine local behavior,
the pressure of each parcel is set dynamically by flows primarily
driven by distant SNe.  The theory of polytropic, compressible
turbulence described by Passot \& V\'azquez-Semadeni (1998; hereafter
PV98) seems able to explain some of our results, with careful
application.

Gazol et al.\ (2001) demonstrate that turbulence driven by ionization
heating results in nearly half of the gas lying in thermally unstable
regions, in agreement with the observations of Heiles (2001).  Earlier
dynamical models of the ISM have been reviewed by Mac Low (2000).
Models including SN driving were first done in two dimensions by Rosen
\& Bregman (1995), and have since been done by a number of other
groups. V\'azquez-Semadeni, Passot \& Pouquet briefly discussed the
pressure structure in a two-dimensional hydrodynamical model including
only photoionization heating.  In three dimensions, Avillez (2000)
used an adaptive mesh refinement code to compute the evolution of a
$1\times 1\times 20$~kpc vertical section of a galactic disk, while
Korpi et al.\ (1999) used a single grid magnetohydrodynamical (MHD)
code to model an SN-driven galactic dynamo. They briefly discuss the
pressure structure as well.  All of these computations show that the
interactions between SNRs drive turbulent flows throughout the ISM,
and that radiative cooling of compressed regions produces cold, dense
clouds with relatively short lifetimes.  

We here study in substantially more detail than previous workers the
pressure distribution in both hydrodynamical and MHD models of the
SN-driven ISM.  For the hydrodynamical case, we use the same code as
Avillez (2000), including stratification, radiative cooling, and both
clustered and isolated SNe.  For the MHD case, we use the Riemann
framework of Balsara (2000) in single-grid mode on a 200~pc cube with
periodic boundary conditions on all sides, including both radiative
cooling and heating, as well as isolated SNe.  Although these two
models are not directly comparable, they do allow us to examine a wide
range of physical conditions and draw some firm conclusions about the
behavior of the interstellar pressures.

In \S~\ref{sec:models} we describe the different models we use in more
detail.  We then give two different analytic approaches to the
question of the pressure distribution in the SN-driven ISM in
\S~\ref{sec:an-th}, derived from the work of MO and PV98.  In
\S~\ref{sec:numerics} we use the numerical results to demonstrate that
the SN-driven ISM is far from isobaric, with order of magnitude
pressure variations, but that the distribution of pressures takes on a
log-Gaussian form that can be very well described.  These results are
compared to observations in \S~\ref{sec:observations}, and the paper
is summarized in \S~\ref{sec:summary}.

\section{Models}

\label{sec:models}
\subsection{Stratified Gas Dynamical}

The first type of model we use is a three-dimensional computation of
the disk-halo interaction in an SN-driven ISM using an
adaptive-mesh-refinement (AMR) hydrodynamics code on a $1\times
1\times 20$ kpc region of the Galactic disk, described by Avillez
(2000).  The model includes a fixed gravitational field provided by
the stars in the disk, and radiative cooling assuming optically thin
gas in collisional ionization equilibrium.  The radiative cooling
function is a tabulated version of that shown in Figure~2 of Dalgarno
\& McCray (1972) with an ionization fraction of 0.1 at temperatures
below $10^4$~K and a temperature cutoff at 100~K. (No pervasive
heating term is included in this model, however, and so it never
reaches thermal equilibrium.)  The interstellar gas is initially
distributed in a smooth disk with the vertical distribution of the
cool and warm neutral gas given by Lockman, Hobbs, \& Shull (1986) and
summarized in the Dickey \& Lockman (1990) distribution. In addition,
an exponential profile representing the $z-$distribution of the warm
ionized gas with a scale-height of 1 kpc in the Galaxy as described in
Reynolds (1987) is used.

SNe of types Ib, Ic, and II are explicitly set up at random locations
isolated in the field ($40\%$) as well as in locations where previous
SNe occurred, representing OB associations ($60\%$). The latter are
set in a layer of a mean half thickness of 46 pc (from the midplane)
following the distribution of the molecular gas in the Galaxy, while
the isolated SNe are set in a layer having half thickness of 100
pc. The first SNe in associations occur in locations where the local
density is greater than 1 cm$^{-3}$. No density threshold is used to
determine the location where isolated SNe should occur, because their
progenitors drift away from the parental association and therefore,
their site of explosion is not correlated with the local density. The
SNe are set up at the beginning of their Sedov phases, with radii
determined by their progenitor masses. Type II SNe come from early B
stars with masses $7.7 M_{\odot} \le M\le 15 M_{\odot}$ while type Ib
and Ic SNe have progenitors with masses $M\ge 15 M_{\odot}$ (Tammann,
L\"offler, \& Schr\"oder 1994). In our model the maximum mass allowed
for an O star is 30 $M_{\odot}$. Avillez (2000) describes in detail
the algorithm used to set up the isolated and clustered SNe during the
simulations.

These simulations use the piecewise-parabolic method of Colella \&
Woodward (1984), a third-order scheme implemented in a
dimensionally-split (Strange 1968) manner that relies on solutions of
the Riemann problem in each zone rather than on artificial viscosity
to follow shocks.  During the simulation, the mesh is refined
periodically in regions with sharp pressure variations using the AMR
scheme.  The local increase of the number of cells corresponds to an
increase in resolution by a factor of two (that is, every refined cell
is divided into eight new cells). At every new grid the procedure
outlined above is carried out, followed by the correction of fluxes
between the refined and coarse grid cells. The adaptive mesh
refinement scheme is based on Berger \& Colella (1989), but the grid
generation procedure follows that described in Bell et al. (1994).

The computational domain has an area of 1 kpc$^{2}$ and a vertical
extension of 10 kpc on either side of the midplane.  In the
simulations discussed here, AMR is used in the layer
$\left|z\right|\le 500$ pc. In the highest resolution runs, three
levels of refinement are used, yielding a finest resolution of 1.25
pc. For $\left|z \right|> 500$ pc the resolution is 10 pc. Periodic
boundary conditions are used at the vertical boundaries, while outflow
boundary conditions are used at the top and bottom boundaries. The
loss of matter through the upper and lower boundaries after 1 Gyr of
simulation amounts to $12\%$ of the total initial mass used at the
beginning of the simulations (Avillez 2000).

The rates of occurrence of SNe types Ib, Ic in the Galaxy are $2\times
10^{-3}$ yr$^{-1}$, while those of type~II occur at $1.2\times
10^{-2}$ yr$^{-1}$ (Cappellaro et al. 1997). The total rate of these
SNe in the Galaxy is $\tau_{\rm gal} = 1.4\times 10^{-2}$ yr$^{-1}$,
corresponding to a rate of one SN every 71 yr. These rates are
normalized to the volumes of the stellar disk used in the simulations.
In the current work we report simulations, as described in
Table~\ref{tab:runs}, using three SN rates: $\tau / \tau_{\rm gal} =
1$, 6, and~10.  For each rate of SNe we run two simulations with
different finest resolutions of the AMR hierarchy: 1.25 and 2.5 pc.
In the cases reported here, we ran the simulations for 200 Myr, long
enough to establish the disk-halo circulation and reach a steady state
in the simulated thick gas disk.

\subsection{Magnetohydrodynamical}

The MHD calculations were done using the RIEMANN framework for
computational astrophysics, which is based on higher-order Godunov
schemes for MHD (Roe and Balsara 1996; Balsara 1998a,b), and
incorporates schemes for pressure positivity (Balsara \& Spicer
1999a), and divergence-free magnetic fields, (Balsara \& Spicer
1999b).  (The framework also includes parallelized adaptive mesh
refinement [Balsara and Norton 2001; Balsara 2001], though that
capability is not used in the present paper).


In the models presented here, we solve the ideal MHD equations
including both radiative cooling and pervasive heating in a (200
pc)$^3$ periodic computational box, mostly using a grid of 128$^3$
cells. We start the simulations with a uniform density of $2.3 \times
10^{-24}$ g cm$^{-3}$, threaded by a uniform magnetic field in the
$x$-direction with strength 5.8 $\mu$G, a factor of roughly two
stronger than that observed in the Milky Way disk.  This very strong
field maximizes the effects of magnetization on the turbulence, and
may be seen as the other extreme from our hydrodynamical models.
Behavior common to both sets of models can be deduced to be fairly
independent of the magnetic field.

For the cooling, we use a tabulated version of the radiative cooling
curve shown in Figure~1 of MacDonald and Bailey (1981), which is based
on the work of Raymond, Cox \& Smith (1976) and Shapiro and Moore
(1976).  (It falls smoothly from temperatures of order $10^5$~K to
$10^2$~K, not incorporating a sharp cutoff at $10^4$~K due to the
turnoff of Ly$\alpha$ cooling.)  In order to prevent the gas from
cooling below zero, we set the lower temperature cutoff for the
cooling at 100~K.  We also include a diffuse heating term to represent
processes such as photoelectric heating by starlight, which we set
constant in both space and time.  We set the heating level such that
the initial equilibrium temperature determined by heating and cooling
balance is 3000~K.  Since the cooling time is usually shorter than the
dynamical time, we adopt implicit time integration for the cooling and
heating terms.

We explode SNe at a rate of one every 0.1 Myr in our box, twelve times
higher than our present Galactic rate, corresponding to a mild
starburst like M82.  The SNe are permitted to explode at random
positions.  To avoid extremely high initial expansion velocities, we
do not allow SNe to explode in regions with density less than
0.1~cm$^{-3}$; however, unlike Korpi et al.\ (1999), this does still
allow SNe to explode within the shells of pre-existing remnants. They
are not focussed into associations, however, unlike in the stratified
models.

Each SN explosion dumps 10$^{51}$ erg thermal energy into a sphere
with radius 5 pc.  The evolution of the system is determined by the
energy input from SN explosions and diffuse heating and the energy
lost by radiative cooling. We follow the simulations to the point
where the total energy of the system, as well as the energy in the
thermal, kinetic and magnetic variables, has reached a
quasi-stationary value for several million years. 

\section{Analytic Theories}
\label{sec:an-th}

\subsection{Non-interacting Supernova Remnants}
\label{sub:snrs}
The two-phase theory of FGH assumed pressure equilibrium throughout
the ISM, with densities and temperatures fully regulated by heating
and cooling processes acting on timescales shorter than the dynamical
timescale of the gas.  The introduction of SN explosions by Cox \&
Smith (1974) and MO required relaxation of the assumption of pressure
equilibrium, at least inside of expanding SNRs.  In particular,
Jenkins et al.\ (1983) emphasized that MO implicitly makes
a prediction of the spectrum of pressure fluctuations expected from
the passage of SNRs expanding in a clumpy medium.

The theory of pressure fluctuations begins from the scaling in MO,
Appendix B, of the pressure of the low-density intercloud medium $P$
as a function of the probability $Q(R)$ of a point being within a
SNR with radius at least $R$,
\begin{equation} \label{qeqn}
P = P_{c} \left\{\begin{array}{ll} 
       (Q/Q_c)^{-9/14}    & \mbox{ for $Q \leq Q_c$} \\
       0.5 (Q/Q_c)^{-0.9} & \mbox{ for $Q > Q_c$},    \end{array} \right.
\end{equation}
where $Q_c$ is the probability of a point being within a SN
remnant large enough for the swept-up shell of intercloud medium to
have cooled, and $P_c$ is the corresponding pressure of such a
remnant.  For typical values in the Milky Way, including a SN
rate $S = 10^{-13}$~pc$^{-3}$~yr$^{-1}$, MO find by balancing a number
of observational considerations that likely values for these
parameters are $Q_c = 1/2$, and $P_{c}/k = 10^{3.67}$~cm$^{-3}$~K (see
their eq.~9).  Equation~\ref{qeqn} can then be inverted
to give the probability $Q(P)$ of a point being within a SN
remnant with pressure at least $P$, as written in Jenkins et al.\
(1983).

The probability $Q(P)$ represents a cumulative distribution.  The
corresponding differential probability distribution function (PDF) is given
by $-dQ(P)$.  In practice, we compute PDFs by constructing a histogram
of pressure values, with bins of finite size $\Delta P$, so the PDF
predicted by MO is 
\begin{equation}
-(\partial Q/\partial P)\Delta P = \left\{\begin{array}{ll}
    \frac{14Q_c}{9P_c}\left(\frac{P}{P_c}\right)^{-23/9}\Delta P 
         & \mbox{for $P>P_c$},\\
    \frac{20Q_c}{9P_c}\left(\frac{P}{P_c}\right)^{-19/9}\Delta P 
         & \mbox{for $P \leq P_c$}.  \end{array} \right.
\end{equation}
This PDF would diverge towards low pressures if it were not limited by
the consideration that the pressure in an isolated SN remnant will not
fall below the ambient pressure $P_0$, so that there must be a lower
cutoff at that value.  (MO give $P_0/k = 10^{3.10}$~cm$^{-3}$~K for
the same parameters given above.)  This model predicts no
pressures below the ambient value $P_0$.

\subsection{Turbulence}
\label{sub:turb}

An alternative approach to predicting the distribution of interstellar
pressure fluctuations can be derived from recent work on properties of
highly compressible turbulence by PV98\footnote{This important
paper suffers from a number of typographical errors produced by a
last-minute switch of notation (Passot, priv. comm.) to distinguish
their scaling parameter $M$ from their rms Mach number $\tilde{M}$,
which we denote as $M_{\rm rms}$.  We here enumerate those we are
aware of: (1) In the paragraph above their eq.~(17), $\tilde{M}$
should be used on every occasion. (2) In the text immediately below
their Figure 3, $\sigma_s = \tilde{M}$, not $M$. (Note however, that
their Figure 3 is indeed labelled with $M$, not $\tilde{M}$.) (3)
There is an extra $M$ in the first expression of the middle line of
their equation 18, which should be simply $u_{\rm rms}/c(s)$. (4) The
first term in the bracketed exponential in their equation (20) should
contain an additional factor of $u_{\rm rms}^2$ in the denominator.},
inspired by computations of isothermal turbulence by Padoan, Nordlund,
\& Jones (1997) and V\'azquez-Semadeni (1994).  PV98 considered the
PDF of density in a turbulent, polytropic gas with pressure $P =
K\rho^{\gamma}$, where $\gamma$ is the polytropic index. By making the
assumption that the density fluctuations are built up by successive
passages of shocks and rarefaction waves (V\'azquez-Semadeni 1994)
that act as a random multiplicative process, they were able to show
that the density distribution ${\cal P}$ is a log-normal when the gas
is isothermal ($\gamma = 1$)
\begin{equation}
{\cal P}(s)ds = \frac{1}{\sigma_s\sqrt{2\pi}} \exp \left[ -
\frac{(s-s_0)^2}{2\sigma_s^2} \right] ds,
\end{equation}
where the variable $s = \ln \rho/\rho_0$, and $\rho_0$ is the mean
density of the region. The variance of the logs of the densities
was found numerically to be 
\begin{equation}\sigma_s = M_{\rm rms},
\end{equation}
where the scaled root mean square (rms) Mach number $M_{\rm rms} =
v_{\rm rms} / c(\rho_0)$, the ratio of the rms velocity to the sound
speed at a density $\rho_0$, derivable from the polytropic law.  By
mass conservation, the shift of the peak given by $s_0 = -0.5
\sigma_s^2$.  In the non-isothermal case ($\gamma \neq 1$), a
density-dependent rescaling allows PV98 to derive a tilted log-normal
form
\begin{equation}
{\cal P}(s;\gamma)ds = C(\gamma) \exp\left[-\frac{s^2}{2M_{\rm rms}^2}
- \alpha(\gamma) s \right] ds,
\end{equation}
where $C(\gamma)$ is a normalization constant such that the integral
over the distribution is unity, and $\alpha$ satisfies the relation
$\alpha(2-\gamma) = 1-\alpha(\gamma)$, but is independent of the
strength of the turbulence.

From this formalism, we can derive the PDF 
in pressure ${\cal P}(P)$ by simply using the polytropic law.
For convenience, we define $x = \log_{10} P$, so that
\begin{equation}
s = (x - x_0)/(\gamma \epsilon),
\end{equation}
where $x_0 = \log_{10} P(\rho_0)$ and $\epsilon = \log_{10} e$.  Then
the isothermal distribution becomes
\begin{equation}\label{eq:pdfiso}
{\cal P}(x)dx = (\gamma\epsilon\sigma_s\sqrt{2\pi})^{-1} \exp\left\{ -
\frac{[ (x- (x_0 + \gamma\epsilon s_0)]^2}{2(\gamma\epsilon)^2\sigma_s^2} \right\} dx,
\end{equation}
and the nonisothermal distribution can be derived from equation~(20)
of PV98 to have the somewhat ungainly form
\begin{equation} \label{eq:pdfx}
{\cal P}(x;\gamma)dx = \frac{C(\gamma)}{\gamma \epsilon} \exp \left[
     -\frac{(x-x_0)^2}{2 (\gamma \epsilon)^2 (v_{\rm rms}/c_0)^2} 
      \exp \left\{\frac{\gamma-1}{\gamma\epsilon}(x-x_0)\right\} 
     -\frac{\alpha(\gamma) (x-x_0)}{\gamma \epsilon}         \right] dx.
\end{equation}
The dispersion of the decimal logs of the pressures is thus predicted in
either case to be 
\begin{equation} \label{eq:sigma}
\sigma_x^2 = \gamma^2\epsilon^2 (v_{\rm rms}/c_0)^2.
\end{equation}
In equation~(\ref{eq:pdfx}) the exponential factor multiplying the
Gaussian $x^2$ term cuts it off on one side of the peak, allowing the
dominance of the power-law term with slope given by
$\alpha(\gamma)/\gamma\epsilon$, which was numerically found by PV98
to have the value 0.43 for $\gamma = 1.5$.  Below we will compare our
numerical results to this formalism.

\section{Numerical Results}
\label{sec:numerics}

\subsection{Morphology}

\begin{figure}   
\centerline{\hbox{
\psfig{file=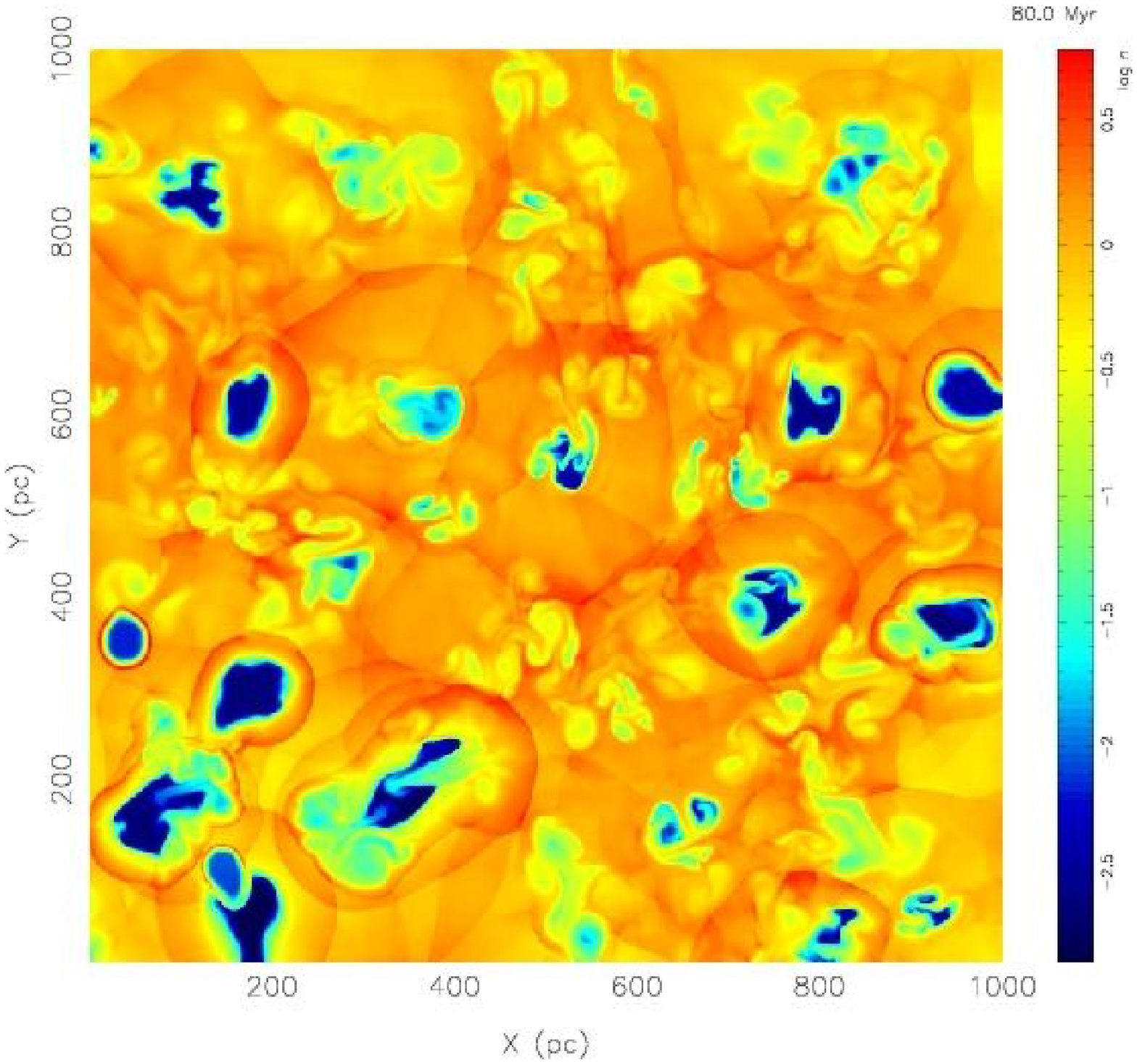,width=0.5\textwidth,angle=270}
\psfig{file=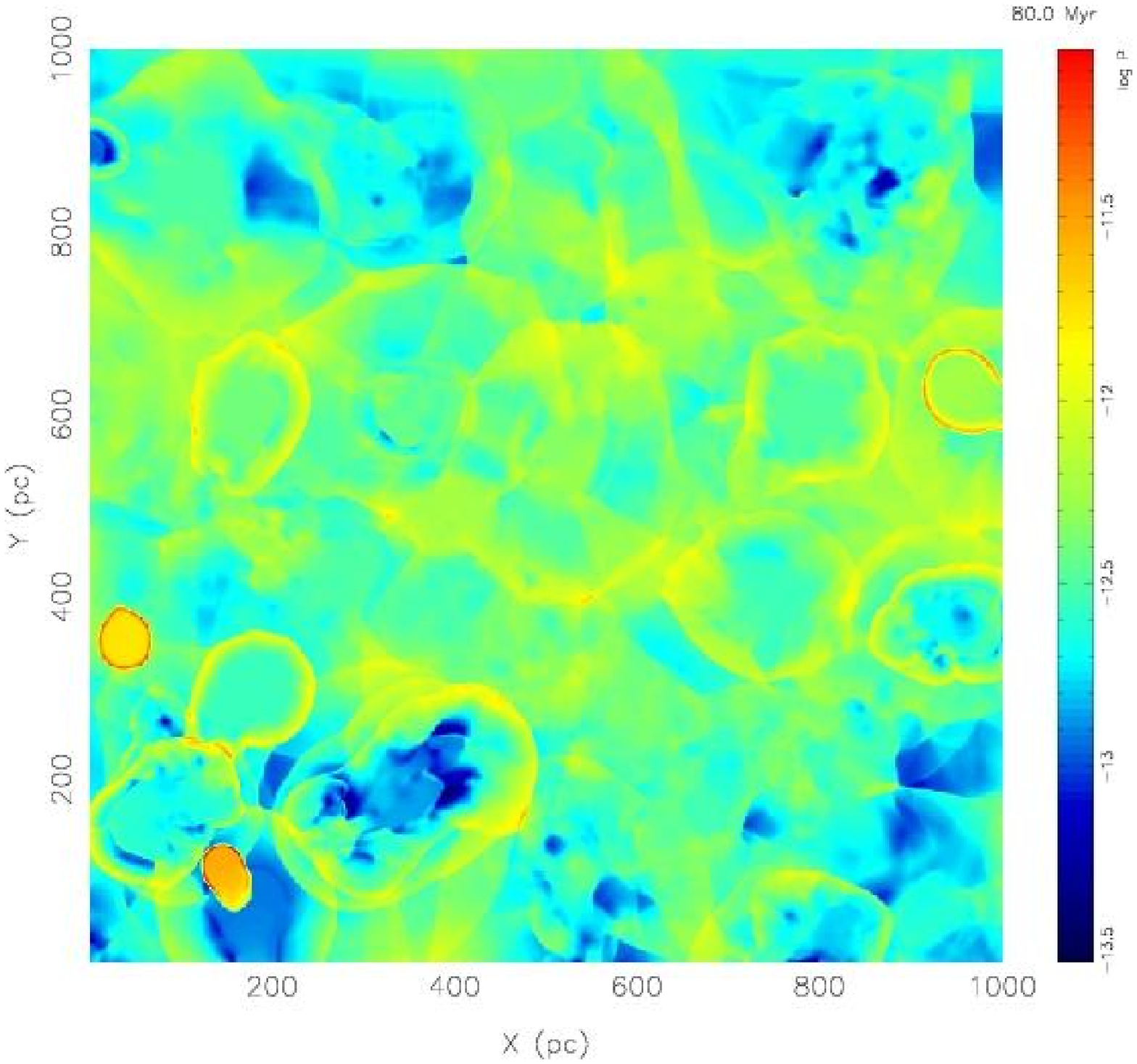,width=0.5\textwidth,angle=270}
}}
\centerline{
\psfig{file=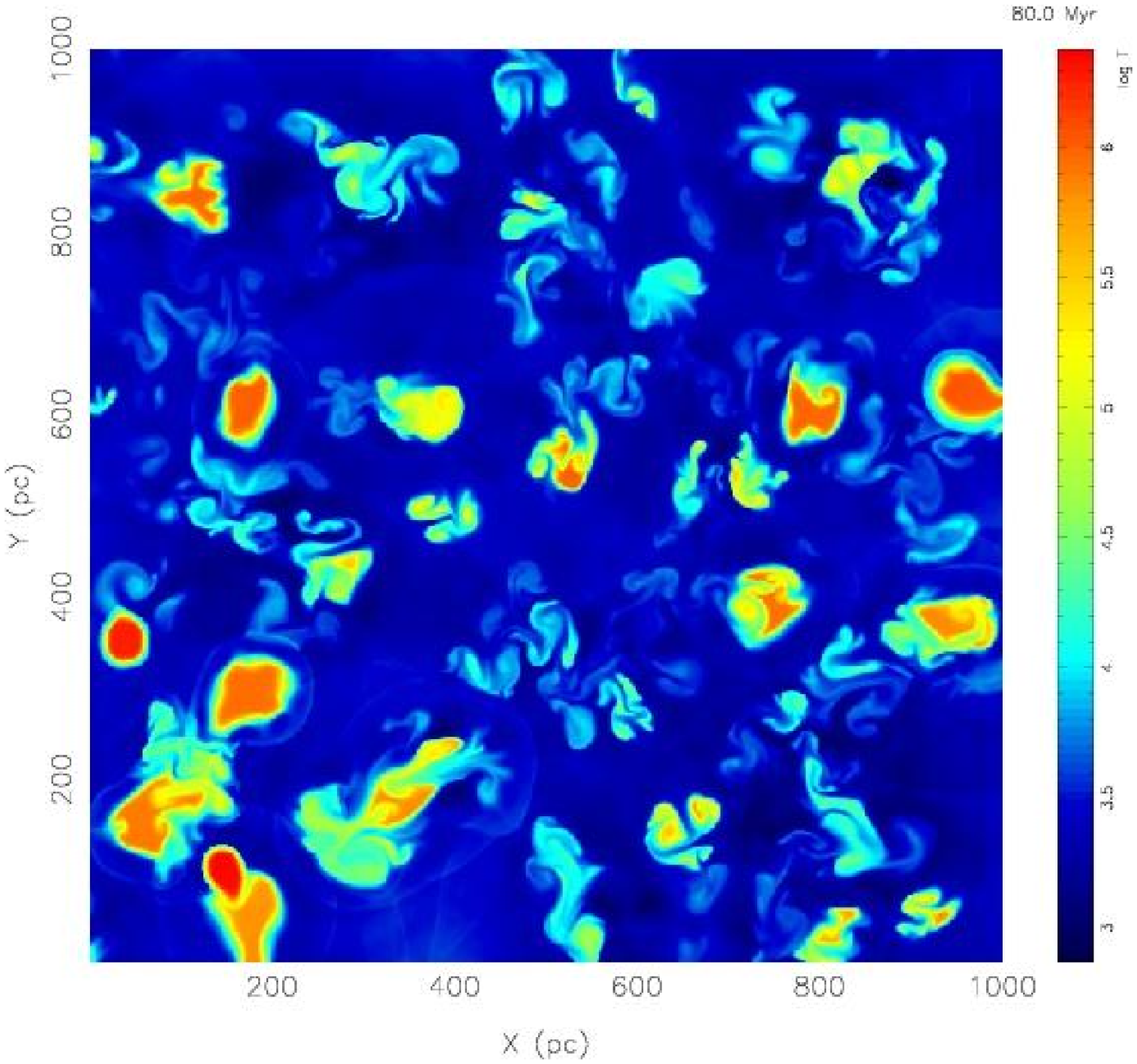,width=0.5\textwidth,angle=270}
}
\caption{\label{fig:strat-cut} Two-dimensional slices through the
three-dimensional stratified model S2 in the Galactic plane at a time
of 80 Myr showing (a) density, (b) thermal pressure,
and (c) temperature. Color bars indicate the scale of each quantity.}
\end{figure}

\begin{figure} 
\centerline{\hbox{
\psfig{file=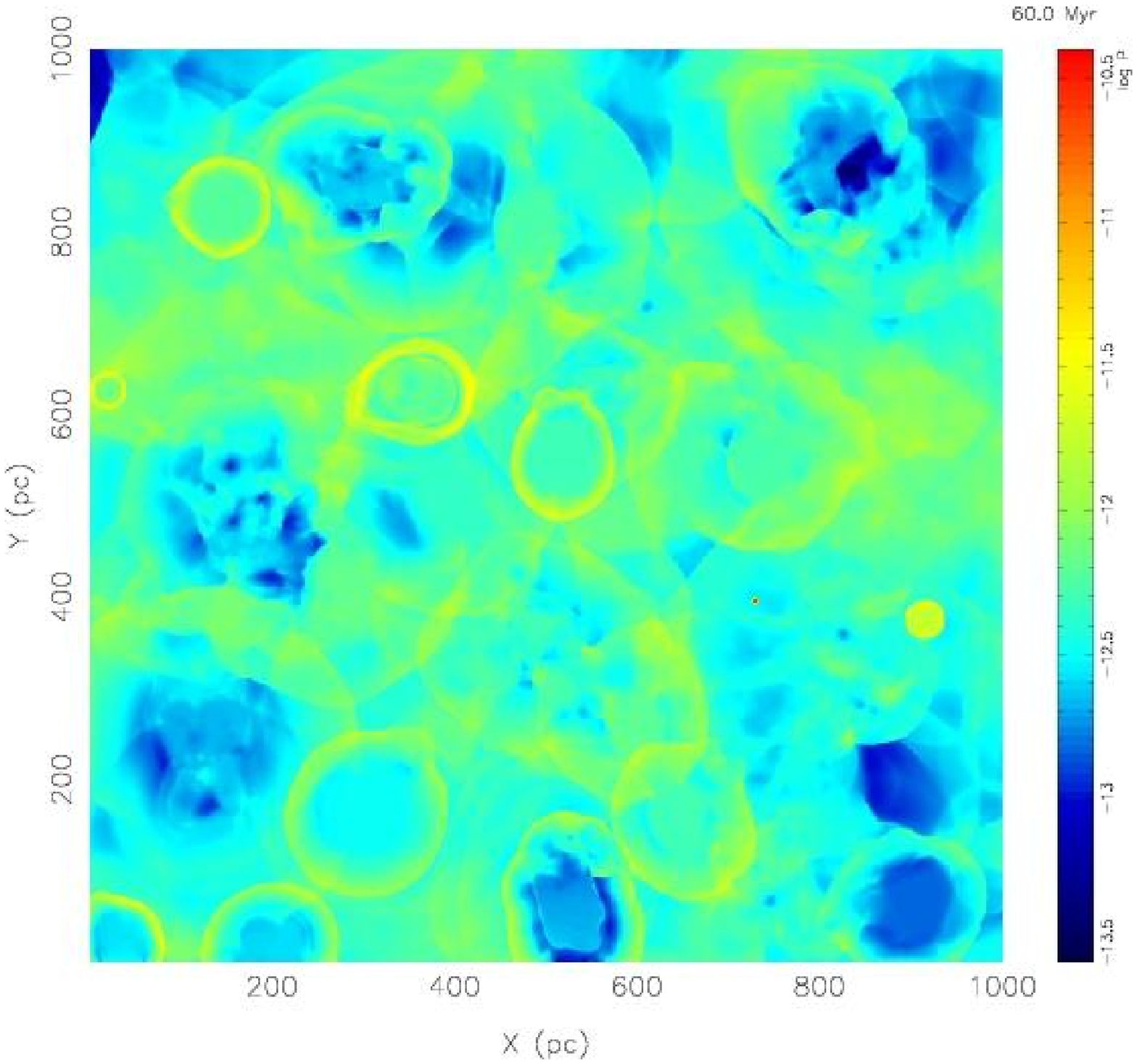,width=0.5\textwidth,angle=270}
\psfig{file=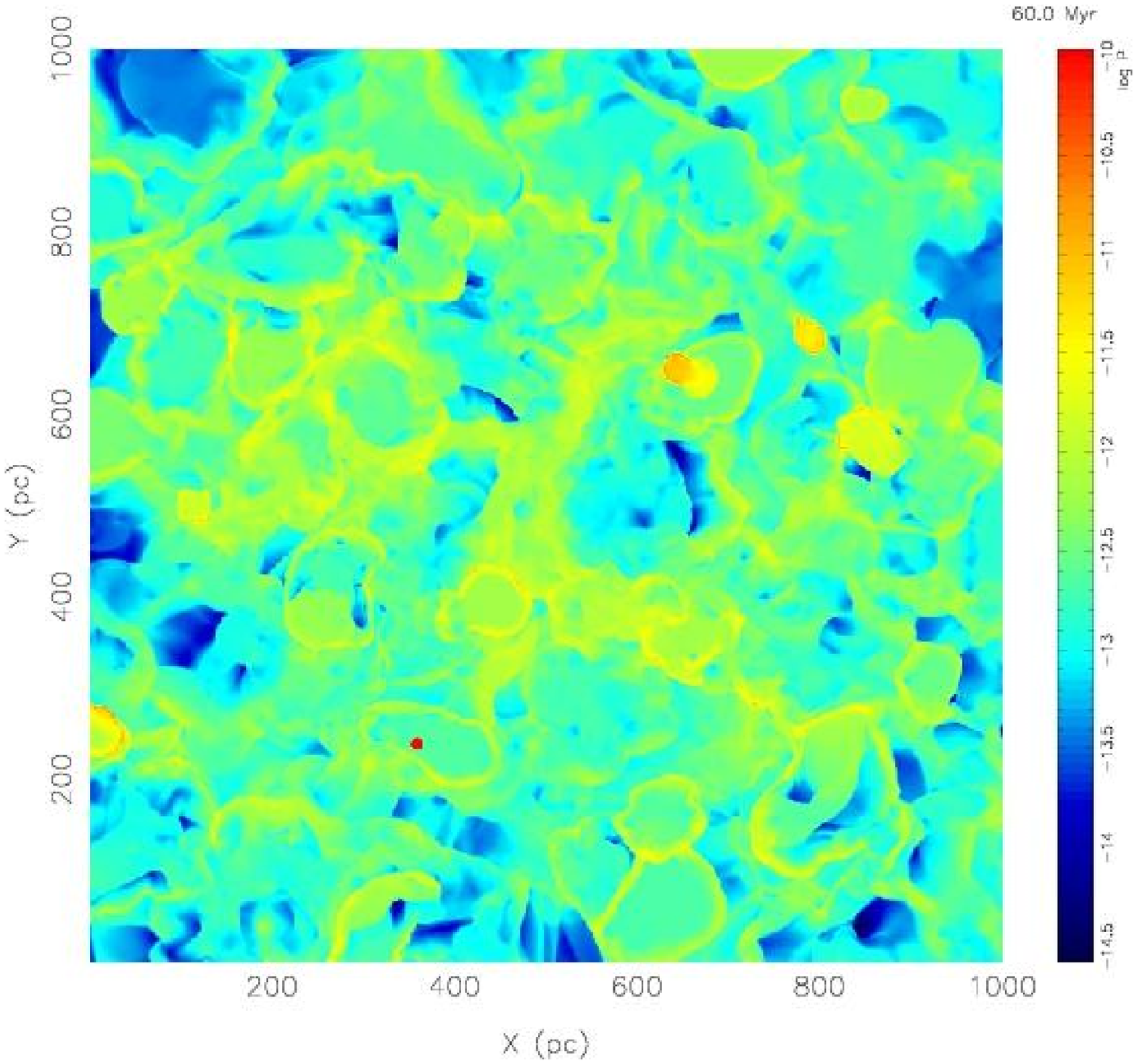,width=0.5\textwidth,angle=270}
}}
\centerline{
\psfig{file=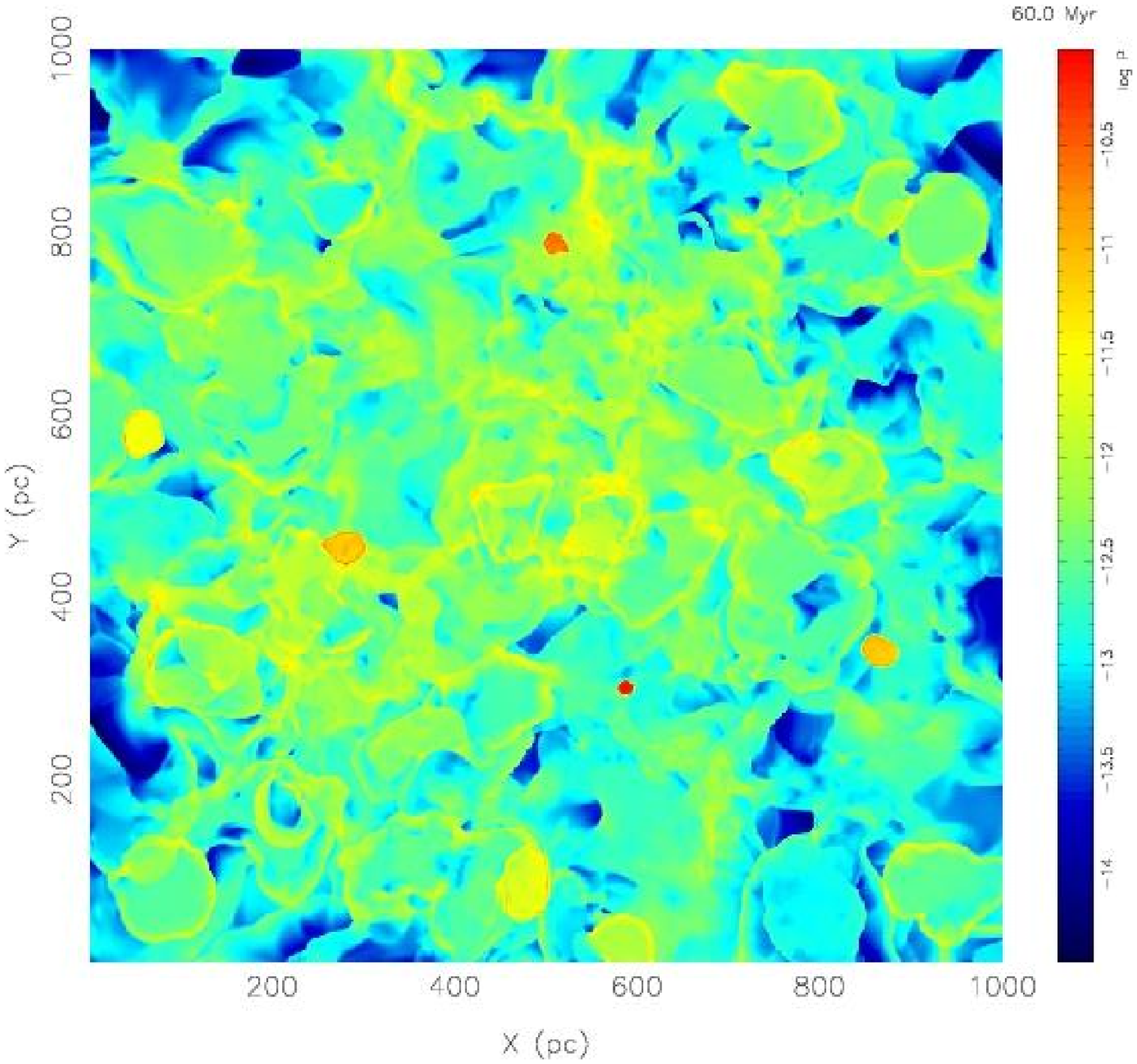,width=0.5\textwidth,angle=270}
}
\caption{\label{fig:strat-cut-rate} Two-dimensional slices through 
three-dimensional stratified models in the Galactic plane at a time
of 60 Myr (20 Myr earlier than in Figure~\ref{fig:strat-cut}) showing
pressure for SN rates of (a) the galactic rate (model S2), (b) six times that
rate ( model S3), and (c) ten times that rate (model S4). 
Color bars indicate the scale of pressure.}
\end{figure}

\begin{figure} 
\centerline{\hbox{
\psfig{file=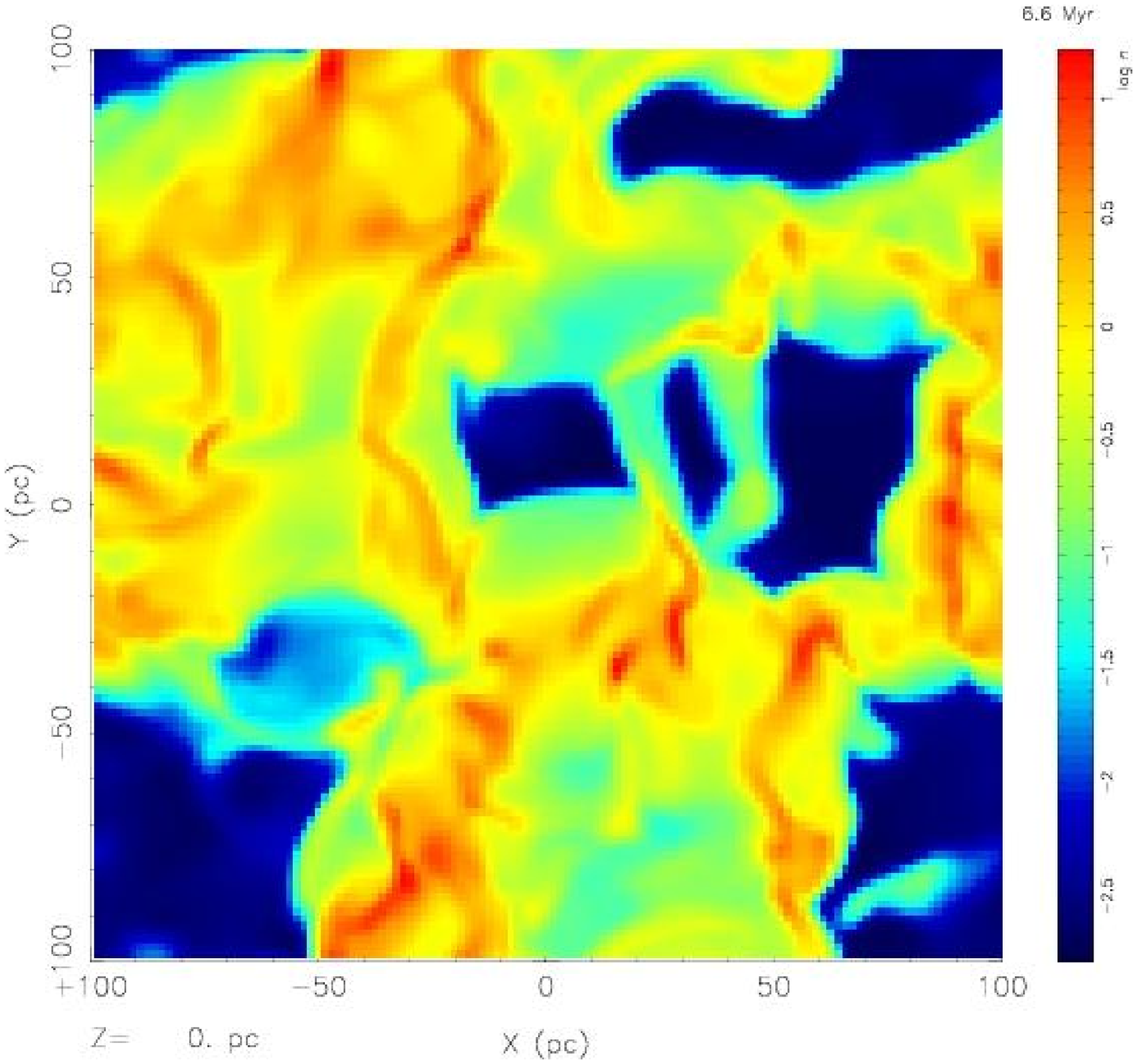,width=0.5\textwidth,angle=270}
\psfig{file=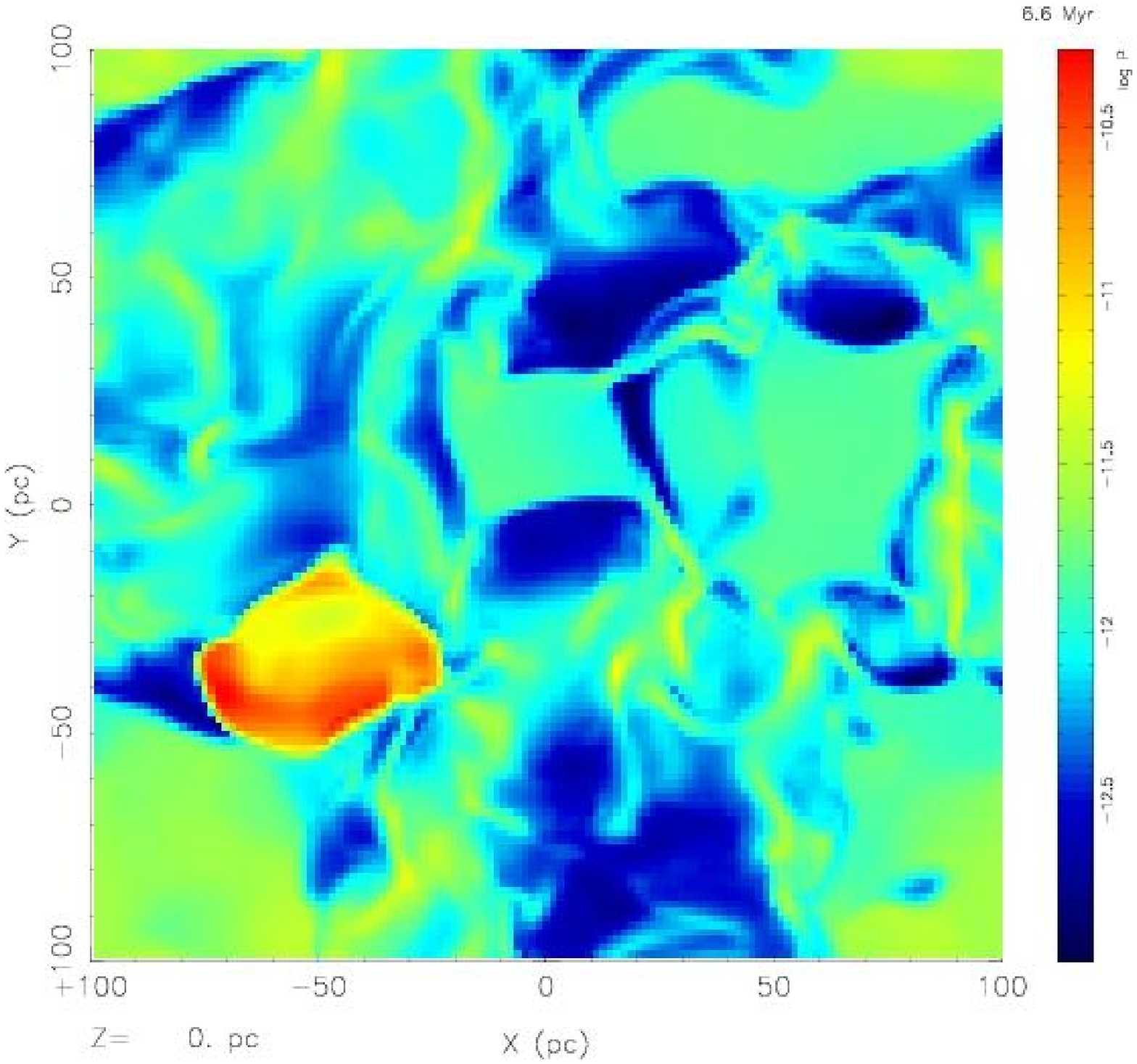,width=0.5\textwidth,angle=270}
}}  
\centerline{\hbox{
\psfig{file=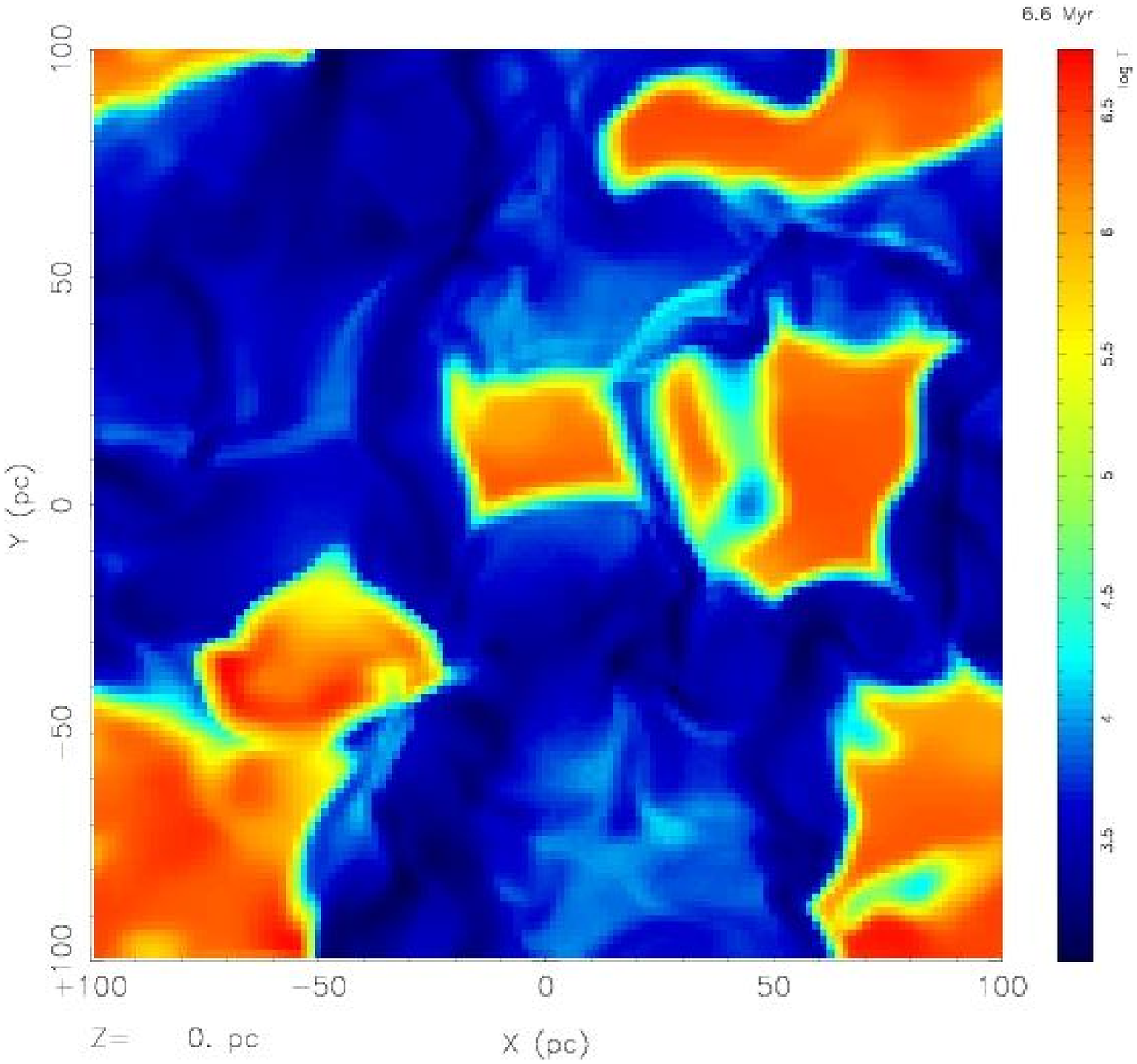,width=0.5\textwidth,angle=270}
\psfig{file=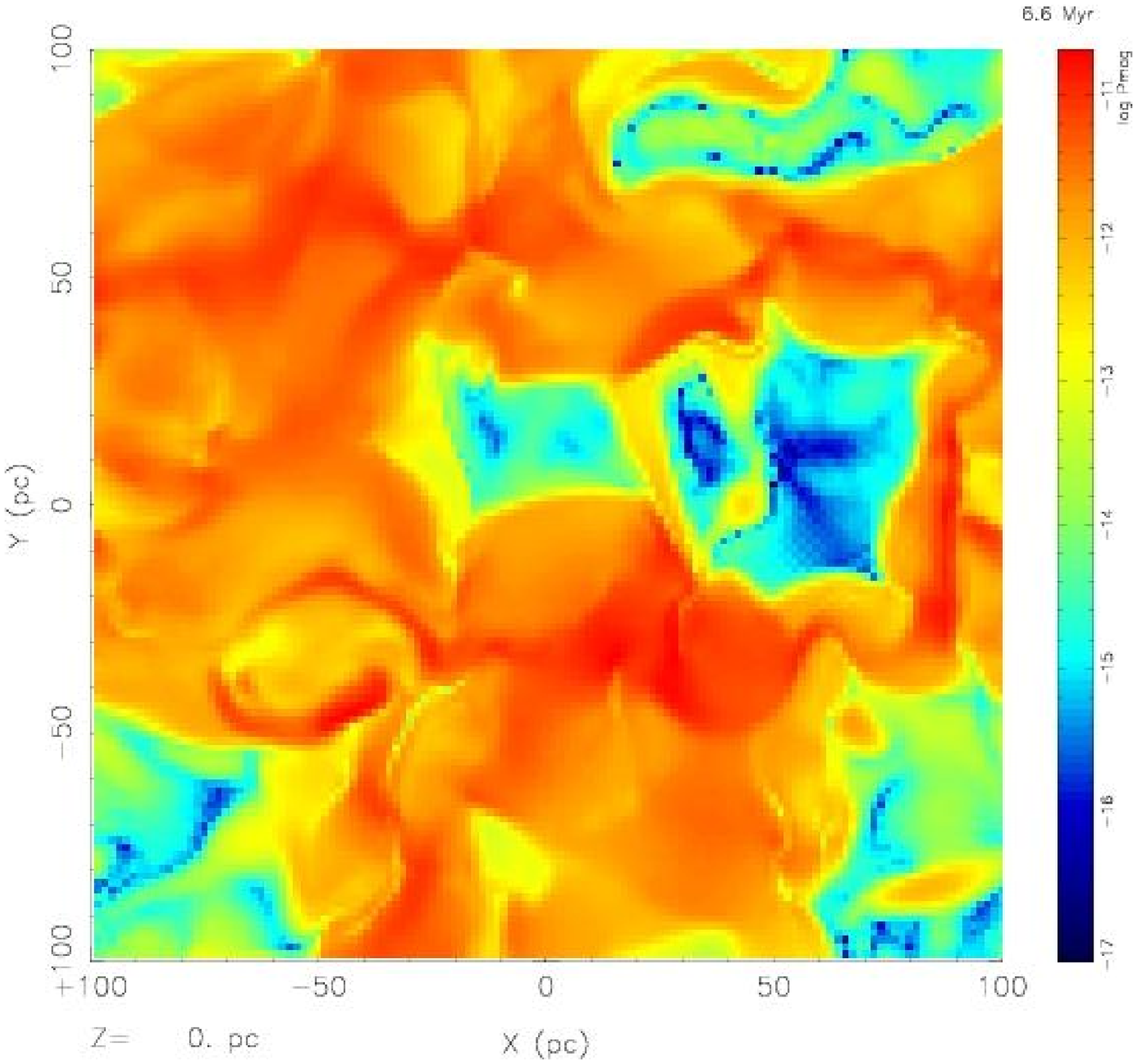,width=0.5\textwidth,angle=270}
}}  
\caption{\label{fig:mhd-cut} Two-dimensional slices through the
three-dimensional MHD model M2, parallel to the magnetic field at a
time of 6.6 Myr, showing density (upper left), thermal pressure
(upper right), temperature (lower left), and magnetic pressure (lower
right). Color bars indicate the scale of each quantity.
}
\end{figure}

We now examine the pressure distribution in our numerical simulations
of a SN-driven interstellar medium.  In Figure~\ref{fig:strat-cut} we
show density, pressure, and temperature on cuts in the plane of the
galaxy from stratified model S2 after it has reached equilibrium, in
Figure~\ref{fig:strat-cut-rate} we compare pressure distributions from
models S2, S3, and S4, with increasing SN rate, and in
Figure~\ref{fig:mhd-cut} we show density, pressure, and temperature,
as well as magnetic pressure, on cuts through the MHD model M2 parallel
to the magnetic field. (The perpendicular direction appears identical,
because the flows are strongly super-Alfv\'enic, with rms Alfv\'en
number exceeding four.) Examination of the pressure images immediately
shows a broad variation in pressures among different regions in all
the models, including in regions not closely associated with young
SNRs.  Regions with pressures markedly lower than ambient are
apparent. 

Low-pressure regions tend to be associated with intermediate density
regions in the hydrodynamical models, while in the magnetized models
the very lowest thermal pressures are actually associated with
substantial magnetic pressures, although there are also low-pressure
regions similar to those in the hydrodynamical models.  We will
examine this more quantitatively further below.  High temperature
regions lie inside young SNRs, while low temperature regions have no
uniform density and temperature correlation.  The highest density
regions have sizes of dozens of parsecs, and average densities
approaching 100~cm$^{-3}$, typical of giant molecular clouds.  This is
consistent with the suggestion by Ballesteros-Paredes et al.\ (1999a)
and Ballesteros-Paredes, V\'azquez-Semadeni \& Scalo (1999) that
molecular clouds are formed and destroyed by the action of the
interstellar turbulence.  Isobaric thermal instabilities, as discussed by
Hennebelle \& P\'erault (1999, 2000), Burkert \& Lin (2000),
V\'azquez-Semadeni, Gazol, \& Scalo (2000), and Gazol et al.\ (2001), must
still be examined, however, by incorporation of appropriate cooling
and heating models at low temperatures. 

Even at SN rates ten or twelve times those characteristic of the Milky
Way, the hot medium in our models does not have the pervasive,
space-filling nature suggested by a simple interpretation of MO.
Rather, discrete regions of hot gas are formed, occasionally
intersect, and then seem to be dynamically mixed back into the warm
gas that fills a substantial fraction of the space.  This large-scale
turbulent mixing, which can most clearly be seen in older remnants in
Figure~\ref{fig:strat-cut}{\em c}, appears to substantially enhance the
cooling rate, while sheets and filaments confined by nearby SNRs seem
more effective at slowing down the expansion of SNRs than the isolated
spherical clouds considered in MO.  However, these results will need
to be confirmed by more careful modeling in the future to ensure that
numerical diffusion is not the main factor in reducing the amount of
hot gas.  The filling factors in the stratified model are discussed in
more detail in Avillez (2000), where it is shown that the filling
factor of the hot gas grows substantially above the disk plane,
ultimately resulting in a galactic fountain above about 1~kpc, even at
SN rates typical of the Milky Way.

\subsection{Thermodynamic Relations}

The first theories of the multi-phase ISM, such as FGH, postulated an
isobaric medium.  Since then, multi-phase models have commonly been
interpreted as being isobaric, although MO and Wolfire et al.\ (1995)
actually assume only local pressure equilibrium, not global, and MO
considered the distribution of pressures, as described above in
\S~\ref{sub:snrs}.  In multi-phase models, the heating and cooling
rates of the gas have different dependences on the temperature and
density, so that the balance between heating and cooling determines
allowed temperatures and densities for any particular pressure.  This
balance can be shown graphically in a phase diagram, showing, for
example, the allowed densities for any pressure (FGH; for a modern
example, see Fig.~3({\em a}) of Wolfire et al.\ 1995).  

In Figure~\ref{fig:mhd-dscat}, the thermal-equilibrium curve for the
heating and cooling mechanisms included in the MHD models is shown as
a black line. (The stratified models did not include any pervasive
heating term and so have no region of true thermal equilibrium.)  Only
a single phase is predicted at high densities as our cooling curve did
not include the physically-expected unstable region at temperatures of
order $10^3$~K (Wolfire et al.\ 1995).  Thus, if our model produced an
isobaric medium, it would be expected to have a single low-temperature
phase in uniform density given by the point at which the
thermal-equilibrium curve crosses that pressure level.  (Effectively,
we would have the hotter two of the three phases proposed by MO.)

The scattered points in Figure~\ref{fig:mhd-dscat} show the actual
density and pressure of individual zones in the model. Many zones at low
temperature do lie on the thermal equilibrium curve, but scattered all
up and down it at many different pressures and densities, with no
well-defined phase structure.  Furthermore, a substantial fraction of
the gas has not had time to reach thermal equilibrium at all after
dynamical compression.  It appears that pressures are determined
dynamically, and the gas then tries to adjust its density and
temperature to reach thermal equilibrium at that pressure.  Most gas
will land on the thermal equilibrium curve when dynamical times are
long compared to heating and cooling times.  This will still lead to
all points within the range of pressures available along the thermal
equilibrium line being occupied, rather than the appearance of
discrete phases.  Unstable regions along the thermal equilibrium curve
(Gazol et al.\ 2001) and off it will also be populated, as observed by
Heiles (2001), but not as densely, as gas will indeed attempt to heat
or cool to a stable thermal equilibrium at its current pressure.

The range of pressures observed in our simulations is, in fact,
broader than the region shown by Wolfire et al.\ (1995) to be subject
to thermal instability.  Even models that included a proper cooling
curve would produce some gas at pressures incapable of supporting a
classical multi-phase structure.  The mixture of different pressures
would, however, produce gas at both high and low densities, as well as
a smaller fraction of gas at intermediate densities that has not yet
reached thermal equilibrium.  We will discuss the resulting density
and temperature PDFs in upcoming work.

Under what conditions will the dynamical times indeed be long compared
to heating and cooling times?  We can attempt to calculate this for
one of our MHD models by making rough analytical estimates of each.  The
dynamical time is
\begin{equation}
t_{\rm dyn} = L/v_{\rm rms},
\end{equation}
where $L$ is a characteristic length scale, while the cooling time 
\begin{equation}
t_{\rm cool} = E/\dot{E} = kT/ n \Lambda,
\end{equation}
where $E$ is the thermal energy, and $\dot{E} = n^2 \Lambda(T)$ is the
cooling rate as a function of temperature $T$. In model M2, the rms
velocity $v_{\rm rms} = 55 \mbox{ km s}^{-1}$.  If we take typical
dynamical length scales of $L\sim 10$~pc, then $t_{\rm dyn} \simeq
0.2$~Myr.  We tabulate $t_{\rm cool}$ from our cooling curve in
Table~\ref{tab:cool}, normalized to a density of $n=1$~cm$^{-3}$. (We
note that model cooling times in gas at temperatures of $10^3\mbox{ K}
\lesssim T \lesssim 10^4$~K are substantially less than physical values,
as the MacDonald \& Bailey cooling curve does not drop abruptly at
$10^4$~K when Ly$\alpha$ cooling shuts off.  However, the consequence
of this is merely that our model overestimates the amount of gas that
has reached thermal equilibrium: the scatter plot shown in
Figure~\ref{fig:mhd-dscat} should be even more uniformly filled.) At all
temperatures $10^3$~K$ < T < 10^7$~K, we find that $t_{\rm cool} \ll
t_{\rm dyn}$, especially at densities $n>1$~cm$^{-3}$, substantiating
our description of the gas dynamics being more important than local
thermodynamics in determining the thermodynamic properties.

The separation between dynamical and thermal timescales also sheds
light on the study by V\'azquez-Semadeni et al.\ (2000) on the effects
of turbulence on thermal instability.  That study suggested that
turbulence erases the effects of thermal instability on the
interstellar medium.  What is suggested by our models is that
turbulence probably didn't erase the thermal instability entirely, but
that the instability only acted locally, under the conditions set for
it by the larger-scale turbulent flow.  Also, thermal instability
becomes less important in determining the overall distribution of
pressures and temperatures when much of the gas has not even reached
thermal equilibrium.  More gas will lie in thermally stable regions
than thermally unstable ones, but the wide range of available
pressures and the lack of complete thermal equilibrium in many regions
still results in a wide range of properties, despite the nominal
action of the thermal instability.

\begin{figure}  
\centerline{\hbox{
\psfig{file=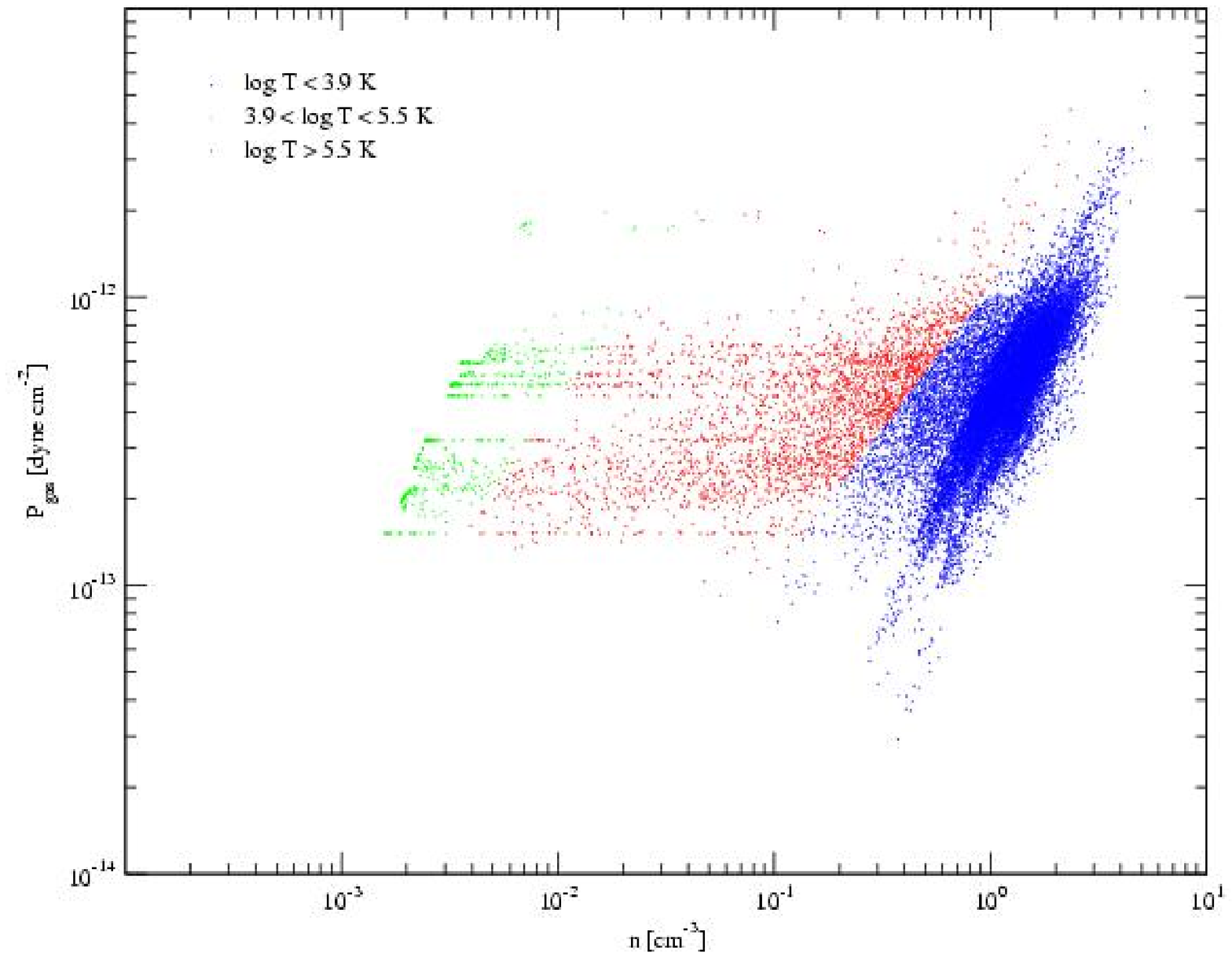,width=0.5\textwidth,angle=270}
\psfig{file=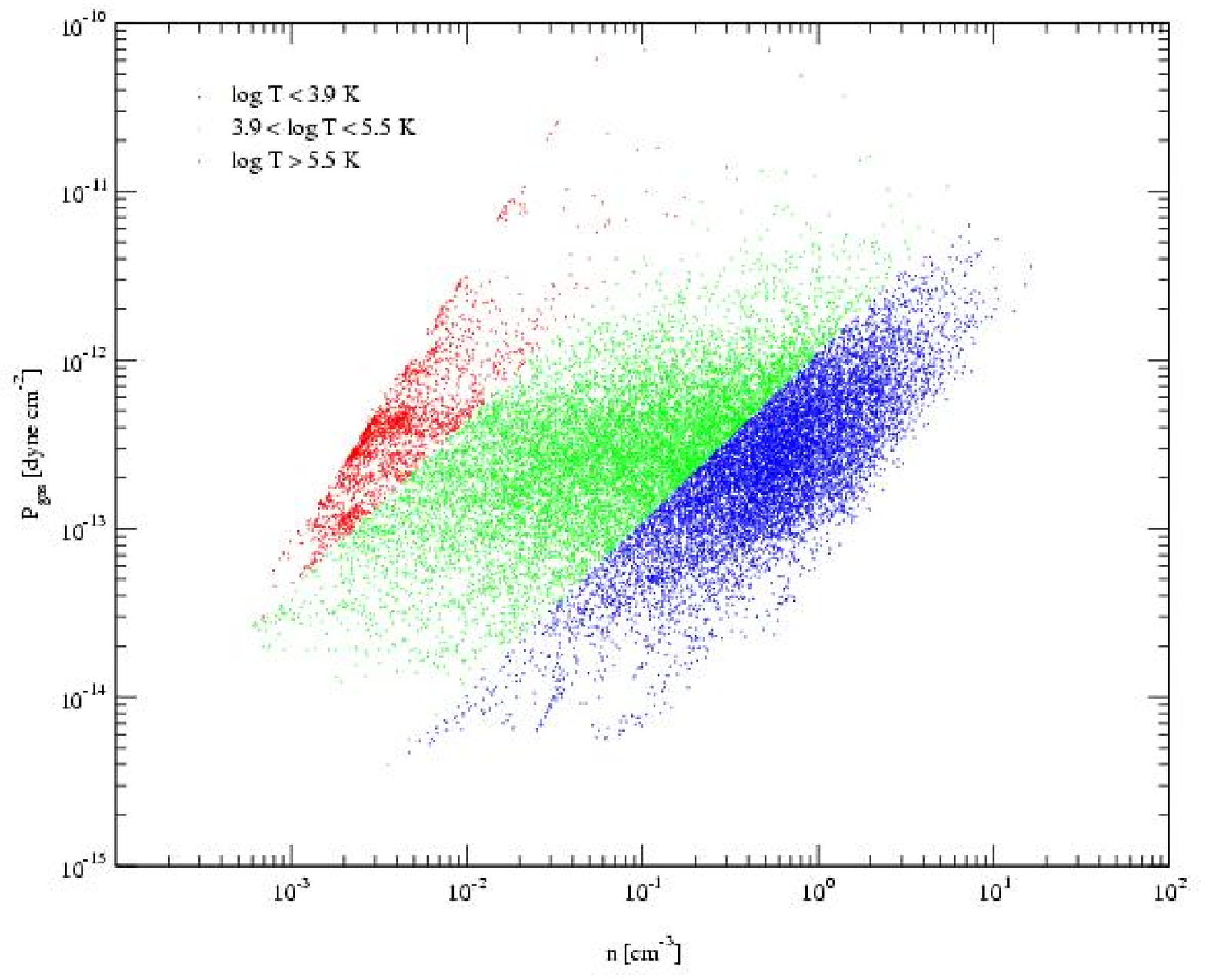,width=0.5\textwidth,angle=270}
}}
\caption{\label{fig:strat-dscat} Scatter plot of pressure vs.\ density
for the stratified models, showing every fourth point in the plane of
the galaxy, with (a) the Galactic SN rate (model S2), and (b) ten
times the Galactic SN rate (model S4).  A wide variation of pressure
is found for each density.  Cool gas with $\log T < 3.9$ is shown in
blue, warm gas with $3.9 < \log T < 5.5$ in green, and hot gas with
$\log T > 5.5$ in red.}
\end{figure}

\begin{figure} 
\psfig{file=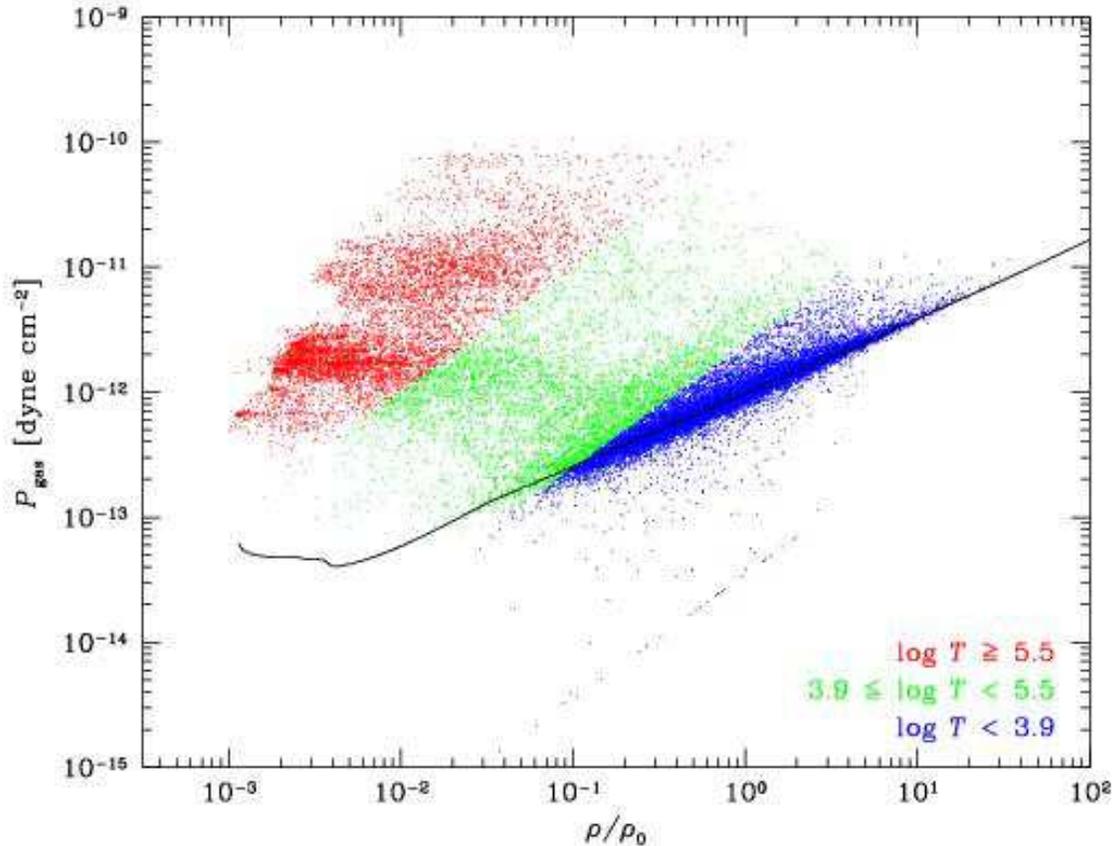,width=\textwidth}
\caption{\label{fig:mhd-dscat}Scatter plot of pressure vs.\ density at
$t=6.6$ Myrs in the MHD simulation M2, showing $32^3$ points sampled
at intervals of four points in each direction.  Note that for each
density a wide variation in pressure is seen. Cool gas with $\log T <
3.9$ is shown in blue, warm gas with $3.9 < \log T < 5.5$ in green,
and hot gas with $\log T > 5.5$ in red.  The thermal equilibrium curve
for the cooling and heating functions in this simulation is overlaid
as a black line. (The line of points at the very bottom right
corresponds to an absolute cutoff in the cooling at 100~K that was
enforced on the temperature in this model.) Note that our cooling
curve may artificially prevent much low-temperature gas from forming
in this model.}
\end{figure}

A separation between dynamical and thermal timescales is also seen in
the stratified hydrodynamical models (Fig.~\ref{fig:strat-dscat}.
However, these did not include an explicit heating term, but rather
had a sharp drop in the cooling rate around $10^4$~K, as found in the
Dalgarno \& McCray (1972) cooling curves.  Below that temperature, the
cooling timescale was longer than the dynamical timescale rather than
shorter, so not much gas had time to cool below it.

The lower boundary of the heavily occupied region on the
pressure-density plane appears to be determined by the polytropic
behavior of the gas near this cutoff.  Fitting to its slope yields
values of the polytropic index $\gamma \sim$~0.6--0.7 in the
magnetized model, where a cooling curve increasing as a power law
(roughly $T^{2.9}$ can be fit) is balanced by heating, while the slope
is much closer to $\gamma \sim $5/3 in the stratified model, where
cooling drops fairly abruptly at a certain temperature, and the bulk
of the gas at that temperature behaves adiabatically.

Points lying below and to the right of this cutoff line at low
pressures and high densities appear to be governed by two different
physical mechanisms.  In the absence of magnetic fields or heating,
but the presence of a low rate of cooling, gas that has been
undisturbed for long enough continues to slowly cool, drifting to the
right of the line.  These points are seen in
Figure~\ref{fig:strat-dscat}.  Low pressures to the left of the line
where cooling cuts off in the stratified model also appear, at low to
intermediate densities of $10^{-2}$--$10^{-1}$~K~cm$^{-3}$.  These are
due to rarefaction waves generated by the turbulent flow acting in gas
that is still cooling or has cooled. In the MHD case, some gas becomes
magnetically supported, dropping to very low thermal pressures at
intermediate densities.  As we will show next, this gas does not have
low total pressure, only low thermal pressure.

\begin{figure} 
\centerline{\hbox{
\psfig{file=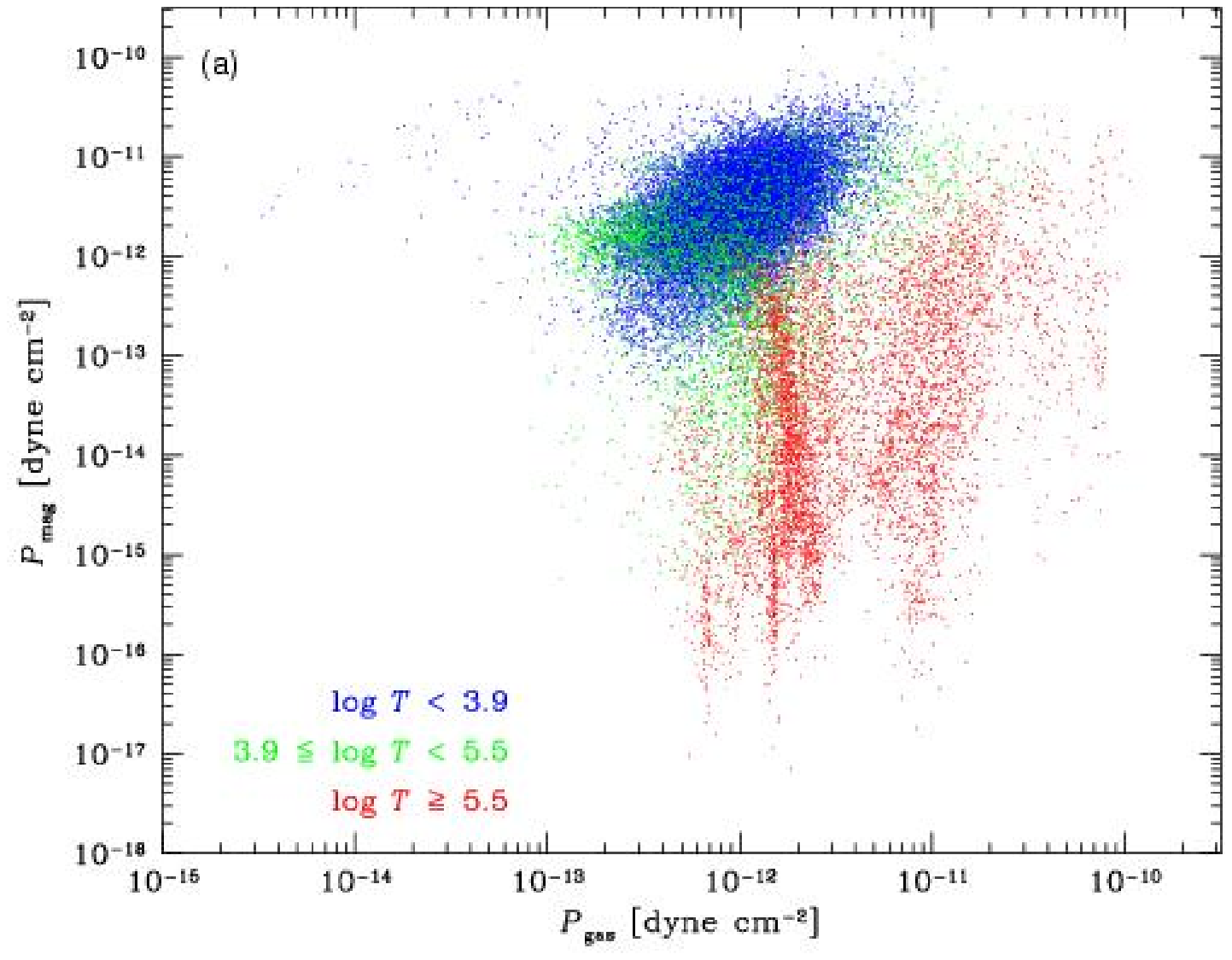,width=0.5\textwidth}
\psfig{file=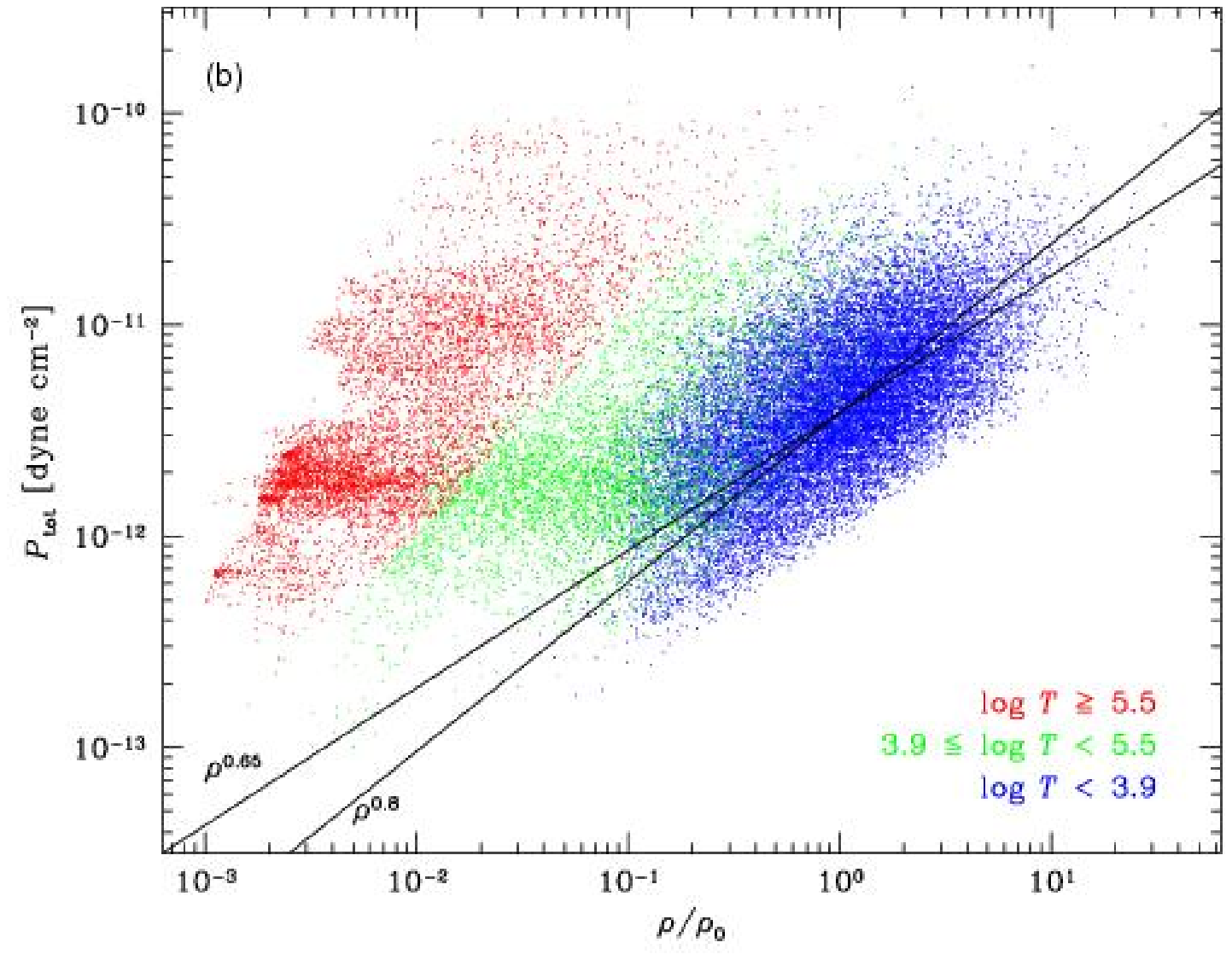,width=0.5\textwidth}
}}
\caption{\label{fig:mhd-mscat} Scatter plot of (a) magnetic vs.\ thermal
pressure and (b) total pressure vs.\ density at $t=6.6$ Myrs
in the MHD simulation, with lines of varying slopes given for
comparison to the lower temperature cutoff.  We again plotted a subset
of $32^3$ points sampled at intervals of four points in each
direction.  Note that regions of very low thermal pressure have
substantial magnetic pressures.  The total pressure vs.\ density
resembles the hydrodynamic results more strongly than the thermal
pressure vs.\ density plots.  
} 
\end{figure}

The relation between magnetic and thermal pressure is shown in
Figure~\ref{fig:mhd-mscat}. In Figure~\ref{fig:mhd-mscat}{\em (a)},
the relative strength of thermal and magnetic pressure is shown at one
time for the magnetized simulation.  The scattering of regions at very
low thermal pressure all have substantial magnetic pressures,
demonstrating that magnetically supported regions can occur.  However,
their relative importance is rather low, as shown by the small number
of points in that regime.  Hot gas can be seen, on the other hand, to
be dominated by thermal pressure, with low magnetic pressures.  In
Figure~\ref{fig:mhd-mscat}{\em (b)}, the total pressure is shown as a
function of density for the same zones. Using total pressure, a clear
cutoff at high densities and low pressures is now plain to see,
showing that the previous scatter of points beyond it was due to the
small magnetically supported regions.

\begin{figure} 
\centerline{\hbox{
\psfig{file=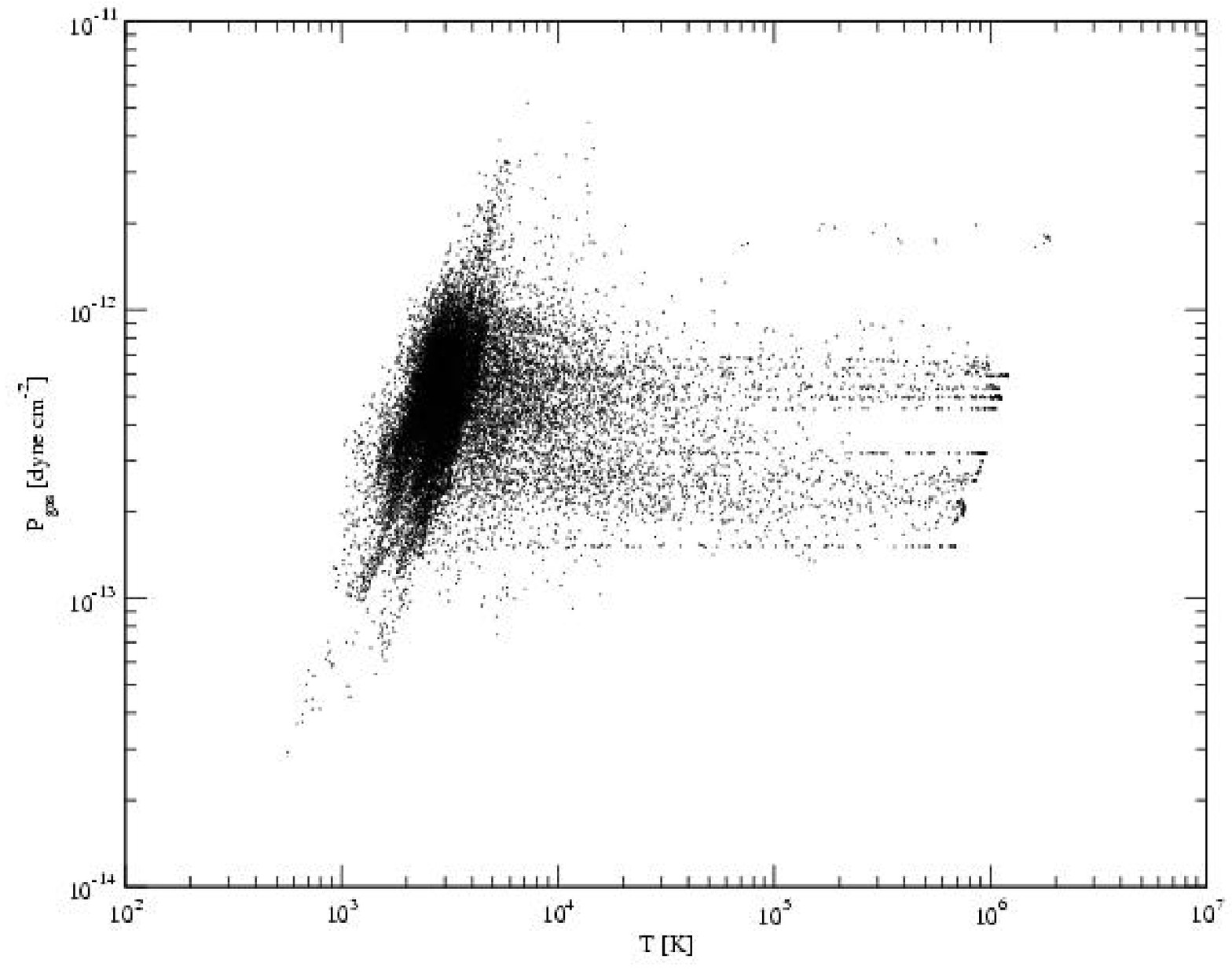,angle=270,width=0.5\textwidth}
\psfig{file=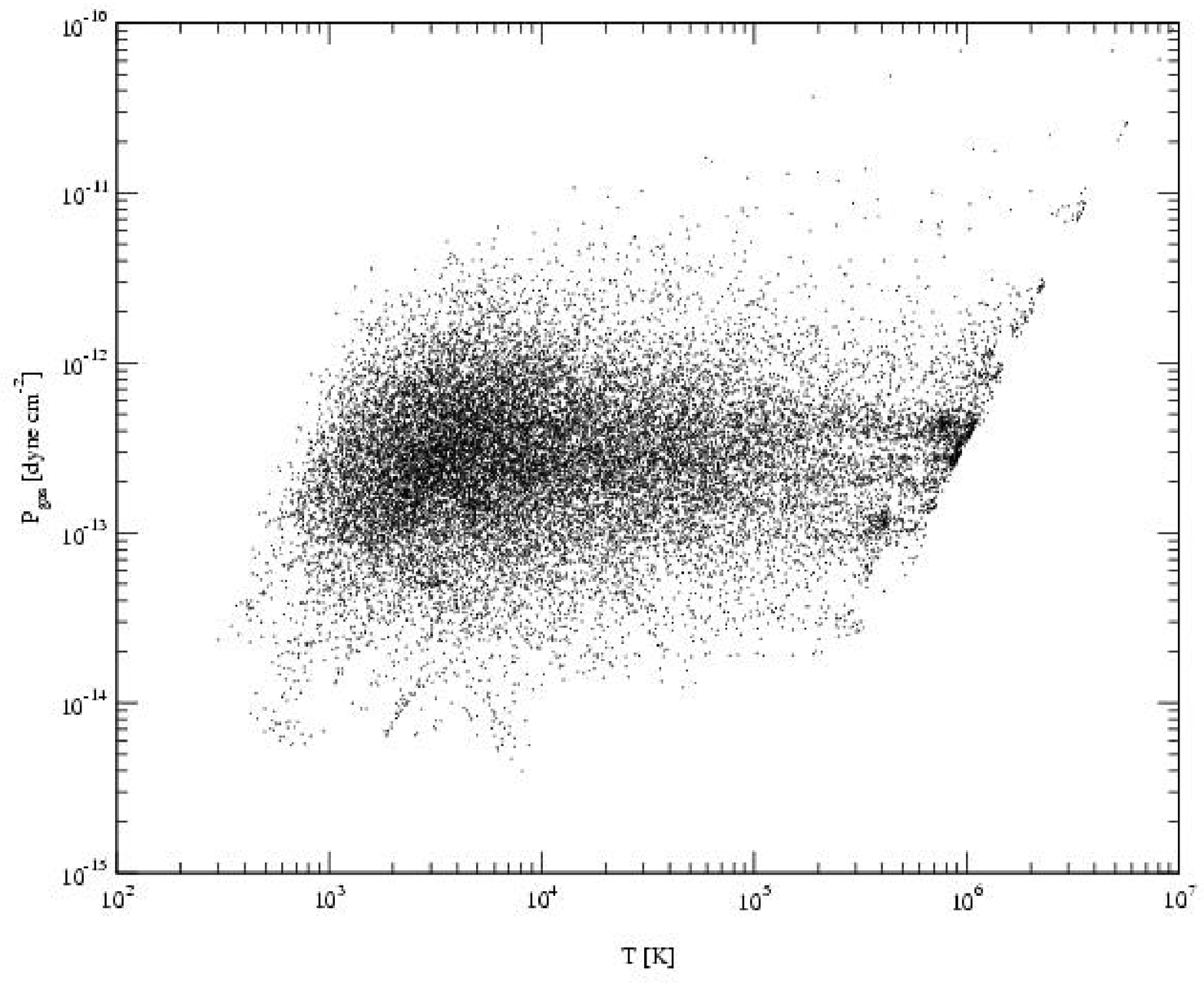,angle=270,width=0.5\textwidth}
}}
\caption{\label{fig:strat-tscat} Scatter plot of pressure vs.\
temperature for a every fourth point in the plane of the galaxy at a
time of 60 Myr for (a) the galactic SN rate (model S2), and (b) ten
times the Galactic rate (model S4).  Note that the pressure scale is
substantially broader in plot (b). }
\end{figure}

\begin{figure}  
\psfig{file=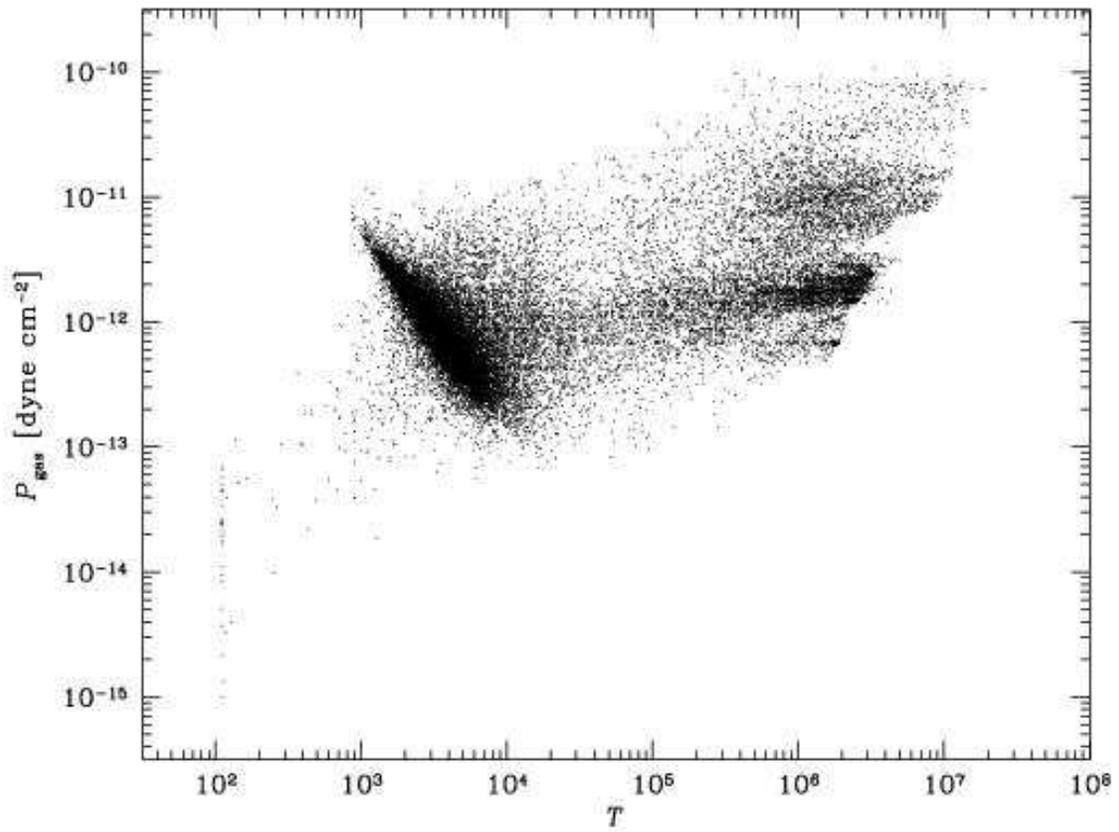,width=\textwidth}
\caption{\label{fig:mhd-tscat}
Scatter plot of pressure vs.\ temperature at $t=6.6$ Myrs in MHD
model M2, again showing a subset of $32^3$ points sampled at intervals of
four points in each direction.  
}
\end{figure}

Finally, let us directly consider the distribution of pressures in gas
at different temperatures.  In Figures~\ref{fig:strat-tscat}
and~\ref{fig:mhd-tscat} we show scatter plots of thermal pressure
against temperature.  We note the concentration of points at
temperatures below $10^4$~K, which once again reflect gas tending to
pile up at the drop in the cooling curve or in thermal equilibrium at
a wide variety of pressures.  The tilt to the right of the cutoff in
Figure~\ref{fig:strat-tscat} again indicates $\gamma > 1$, with gas at
higher pressures also having higher temperatures, while the tilt to
the left in Figure~\ref{fig:mhd-tscat} indicates $\gamma < 1$, with
higher pressure gas typically having lower temperatures.  Individual
SNRs with roughly constant pressures are visible as stripes in the
higher temperature regions of the plots.  As they cool, they also
expand to lower pressures.

\subsection{Model Probability Distribution Functions}

\begin{figure}  
\centerline{\hbox{
\psfig{file=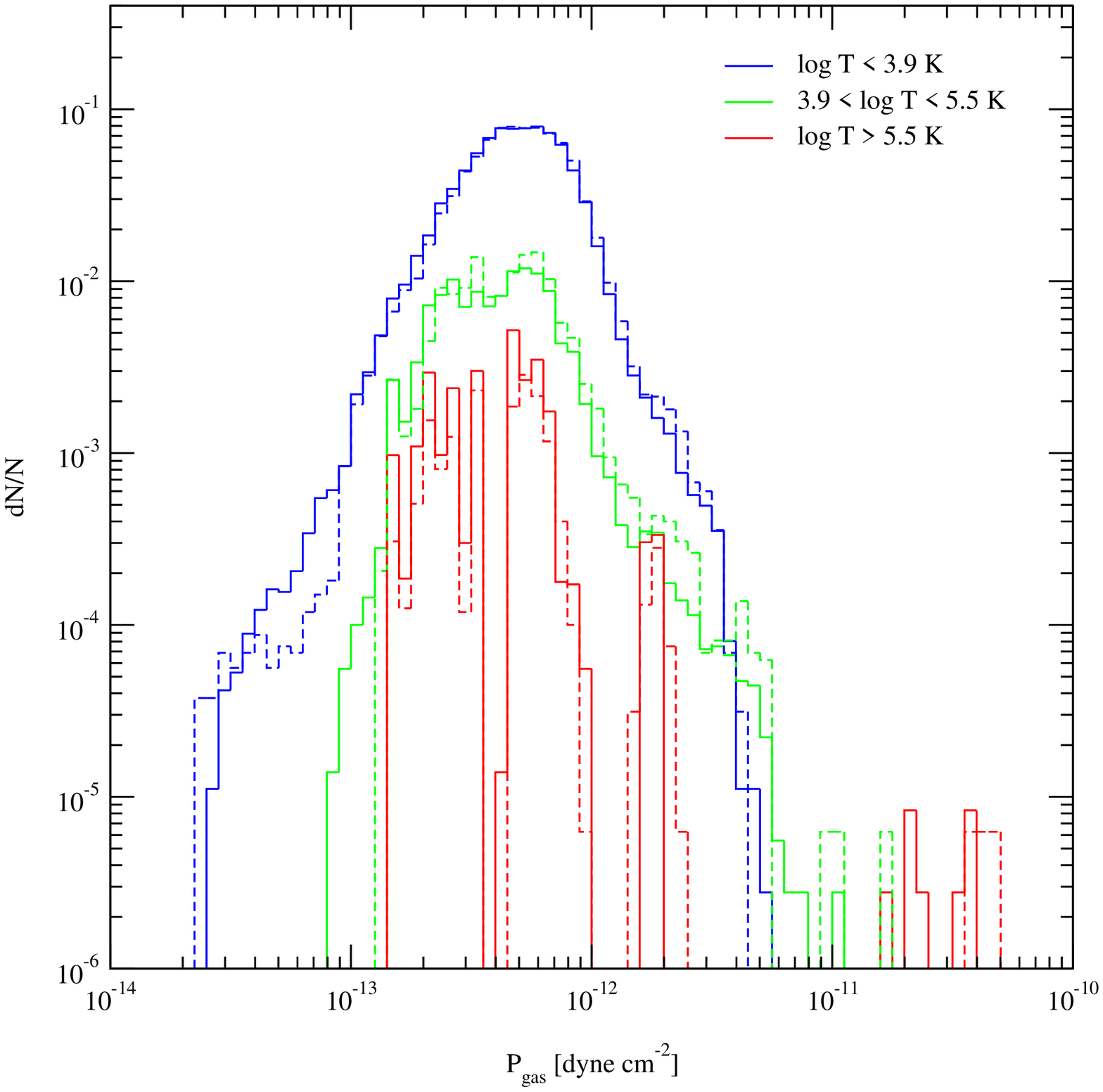,width=0.5\textwidth}
\psfig{file=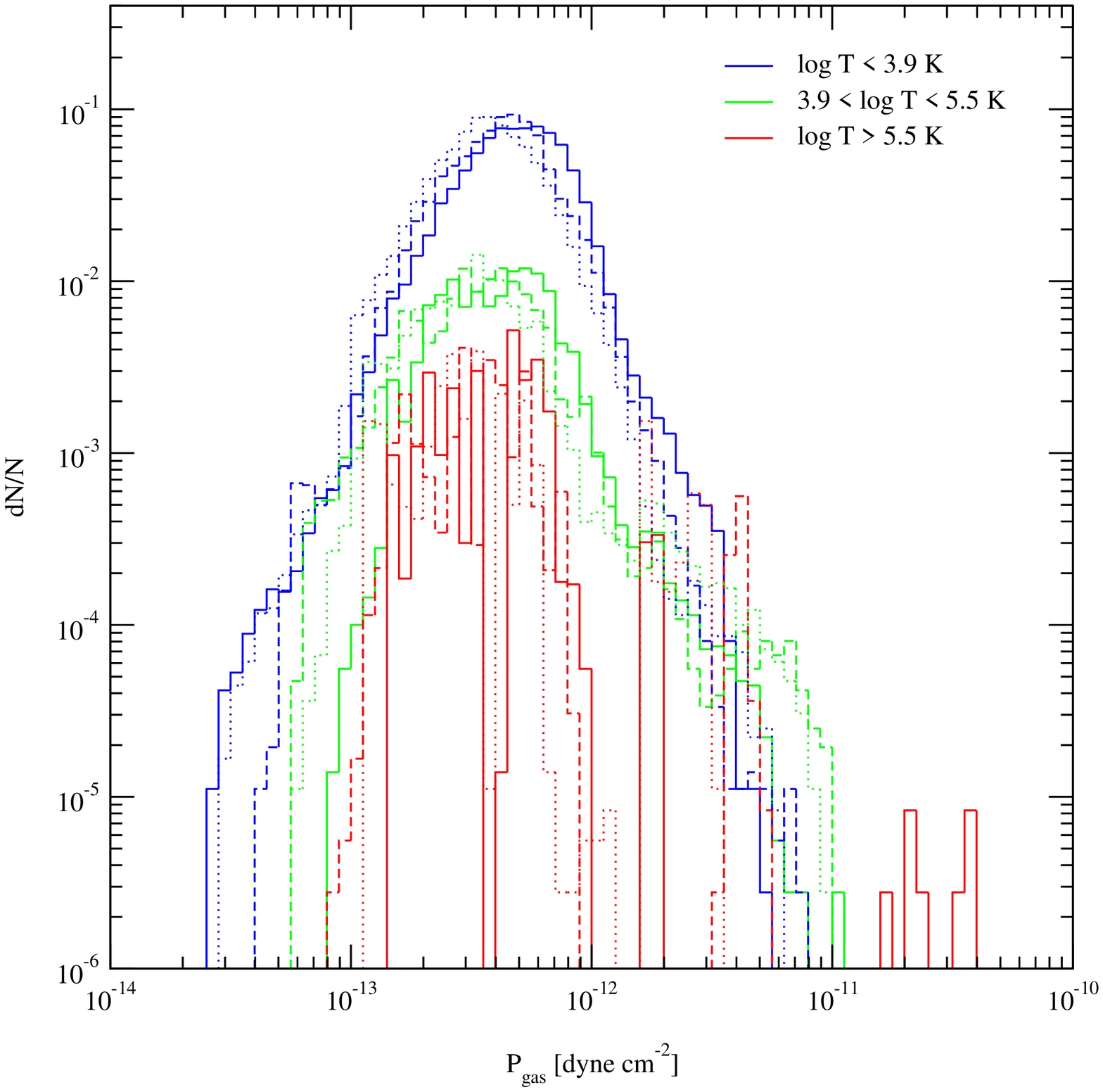,width=0.5\textwidth}
}}
\caption{\label{fig:strat-pdf} Volume-weighted PDFs of pressure from
the stratified models for cool gas with $\log T < 3.9$ (blue), warm
gas with $3.9 < \log T < 5.5$ (green), and hot gas with $\log T > 5.5$
(red) at (a) resolutions of 1.25~pc (solid, model S2) and 2.5~pc
(dashed, model S1) at a time of 60~Myr, and (b) different times of 60
(solid), 70 (dashed), and 80~Myr (dotted) in the 1.25~pc resolution
model S2. This model has a galactic rate of SNe.  }
\end{figure}

\begin{figure}  
\centerline{\hbox{
\psfig{file=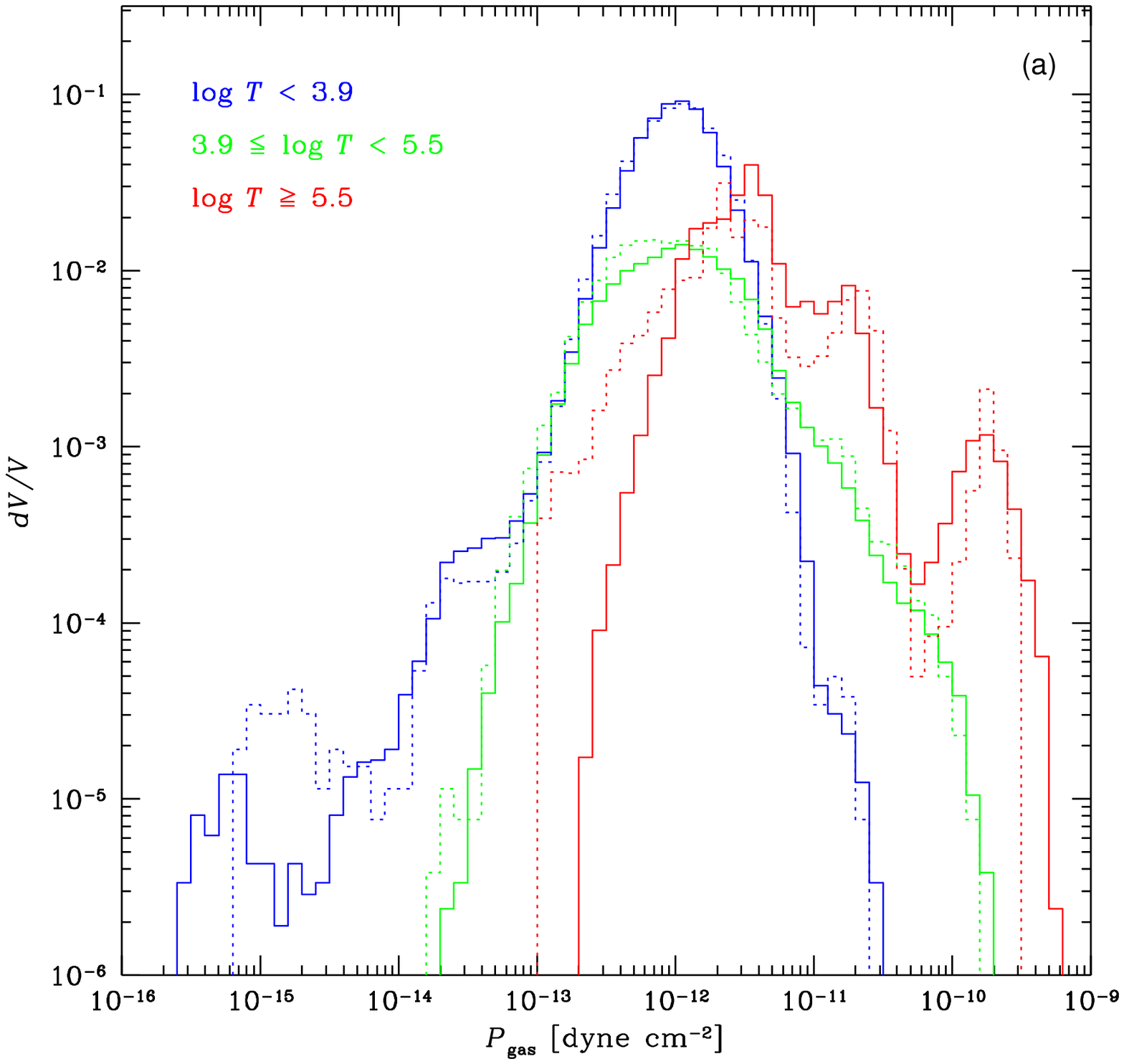,width=0.5\textwidth}
\psfig{file=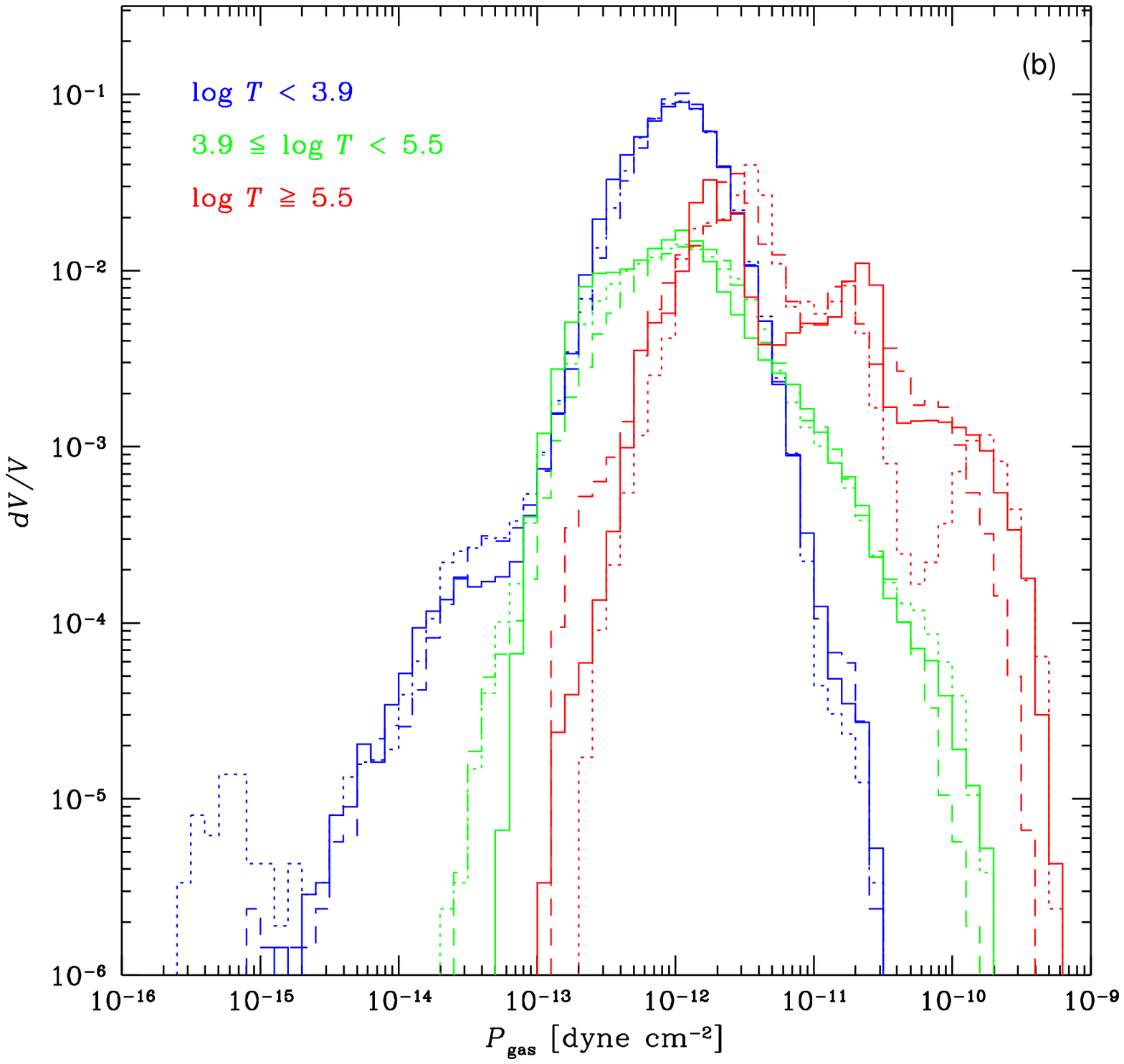,width=0.5\textwidth}
}}
\caption{\label{fig:mhd-pdf} Volume-weighted PDFs of pressure from the
MHD models for cool gas with $\log T < 3.9$ (blue), warm gas with $3.9
< \log T < 5.5$ (green), and hot gas with $\log T > 5.5$ (red) (a) at
different resolutions of 1.6~pc (solid, model M2) and 3.2~pc (dotted,
model M1), at a time of 6.05~Myr, and (b) at different times of
5.55~Myr (dashed), 6.06~Myr (dotted), and 6.55~Myr (solid) in the
1.6~pc resolution model M2. This model has twelve times the galactic rate
of SNe, which results in a much broader pressure distribution in
comparison to Figure~\ref{fig:strat-pdf}, whose $x$-axes cover a
rather shorter range.}
\end{figure}

As we have shown, both the magnetized and the hydrodynamical
models have broad ranges of pressures.  We can quantify this by
examining the pressure PDF, as shown in Figures~\ref{fig:strat-pdf}
and~\ref{fig:mhd-pdf}.  In both cases, these show roughly log-normal
pressure PDFs, very unlike the power-law distributions predicted by
the analytic theory derived by MO, as described in \S~\ref{sub:snrs}.
The observed distributions rather more resemble the pressure
distributions suggested by PV98, as discussed in \S~\ref{sub:turb}.
Not only is the total distribution broad, even at the Galactic~SN rate
(Fig.~\ref{fig:strat-pdf}), but so is the distribution for different
components of the interstellar medium individually.  Furthermore, the
typical or median pressure at the center of the PDF can vary with
temperature, as demonstrated by the PDFs of gas at different
temperature.

Our results do not appear to depend strongly on numerical resolution
as demonstrated in Figures~\ref{fig:strat-pdf}(a)
and~\ref{fig:mhd-pdf}(a) although the details of the history of each
SNR and the total amount of energy radiated away will certainly depend
on the resolution, as well as our neglect of the physics of the
conductive interfaces between hot and warm gas.  

The pressure PDFs also appear to be stable over time, as shown by the
comparison of multiple times in Figures~\ref{fig:strat-pdf}(b)
and~\ref{fig:mhd-pdf}(b), except for the hot gas, especially at the
high-pressure end.  At the high SN rate used in
Figure~\ref{fig:mhd-pdf} individual young SNRs produce discrete bumps
that move left towards lower pressures as time passes, eventually
merging with the overall distribution, while at the lower SN rate used
in Figure~\ref{fig:strat-pdf}, individual SNRs remain distinct for
longer.

\begin{figure} 
\centerline{\hbox{
\psfig{file=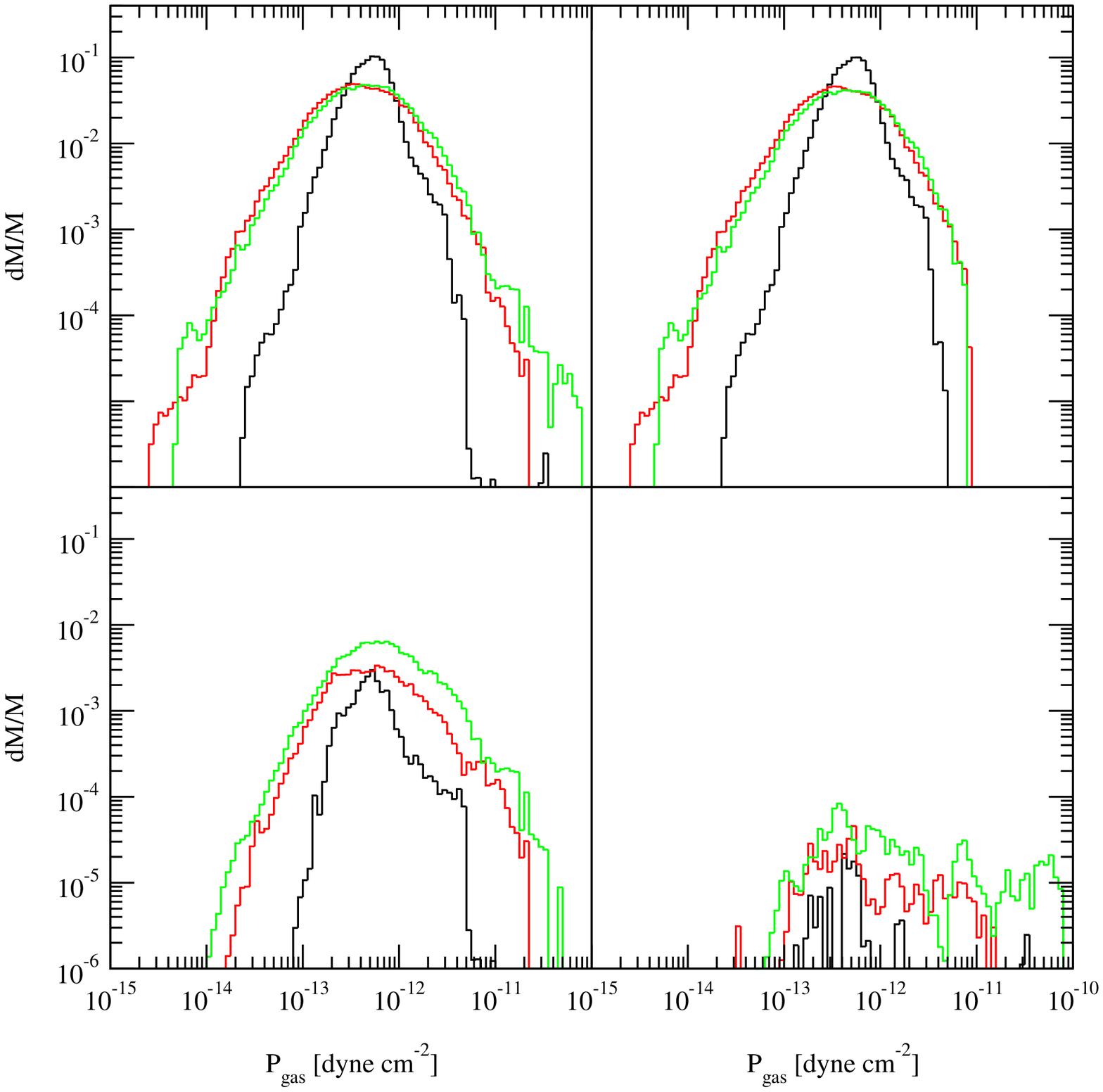,width=0.5\textwidth,angle=0}
\psfig{file=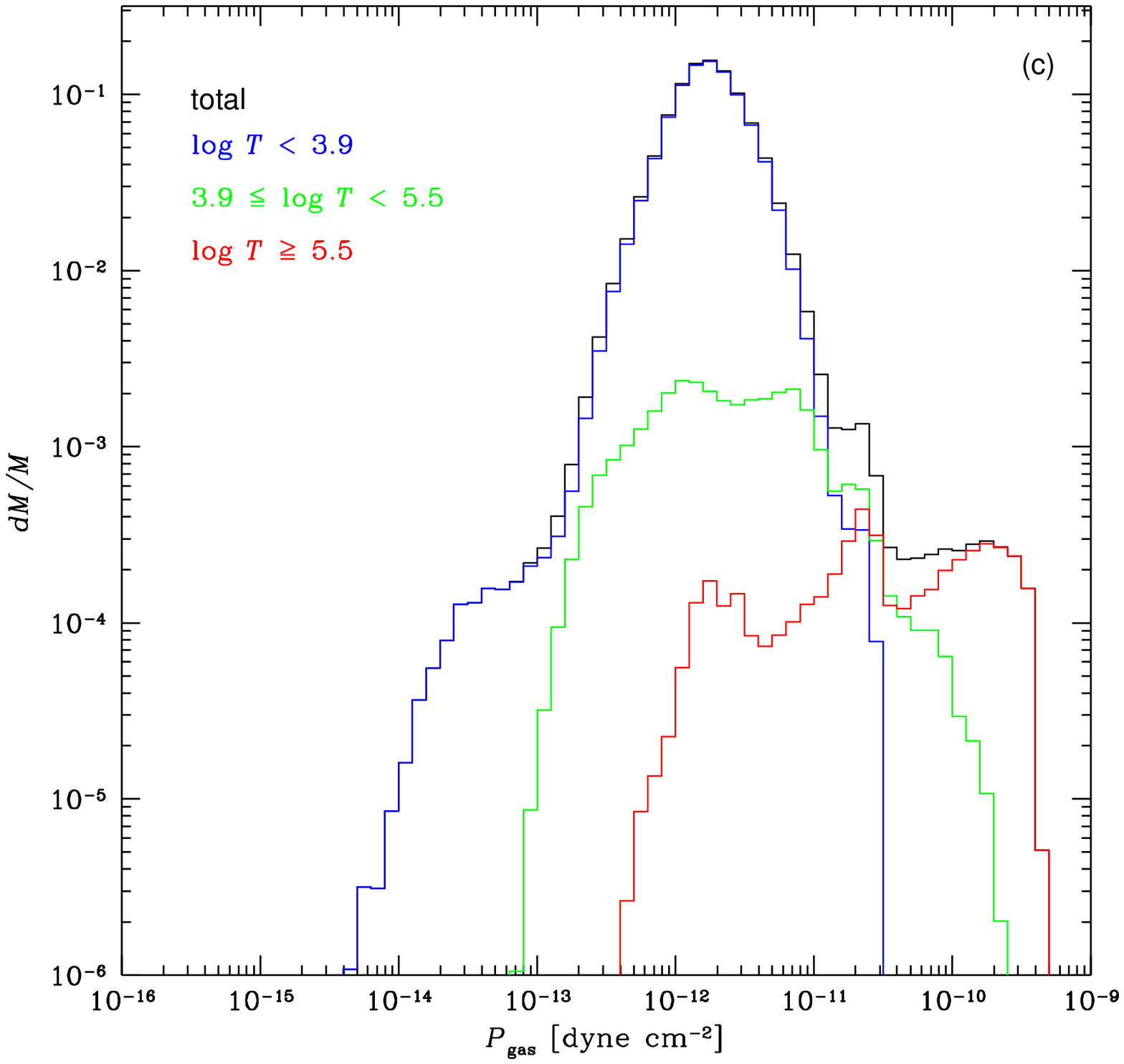,width=0.5\textwidth}
}}
\caption{\label{fig:mass-pdf} Mass distribution of pressure for (a)
the stratified model S1 at a time of 60~Myr for the full distribution
(upper left), and for cool gas with $\log T < 3.9$ (upper right), warm
gas with $3.9 < \log T < 5.5$ (lower left), and hot gas with $\log T >
5.5$ (lower right) at SN rates of 1 (black), 6 (red), and 10 (green)
times the Galactic rate, (b) the MHD model M2 for the full distribution
(black), and for cool gas with $\log T < 3.9$ (blue), warm gas with
$3.9 < \log T < 5.5$ (green), and hot gas with $\log T > 5.5$ (red)
at a time of 6.55~Myrs.  In both models, most of the mass is
found in cold gas, with a broad distribution around the peak pressure.
}
\end{figure}

The question of how mass is distributed among regions of different
pressure becomes important when considering questions such as the
potential ram-pressure confinement of molecular clouds.  The mass
distribution might be expected to markedly diverge from the volume
distribution given by the PDFs shown to date, as most of the hot gas
resides at very low densities.  In Figure~\ref{fig:mass-pdf} we show
mass-weighted distribution functions from each high-resolution model.
The cold gas dominates the mass-weighted PDF, but it is found at the
same wide range of pressures, with roughly the same peak pressure, as
was suggested by the volume-weighted PDFs shown earlier.  In
particular, a substantial fraction of the mass in the cold gas lies at
pressures five to ten times higher than the average pressure, even in
the absence of self-gravity.

\begin{figure}  
\psfig{file=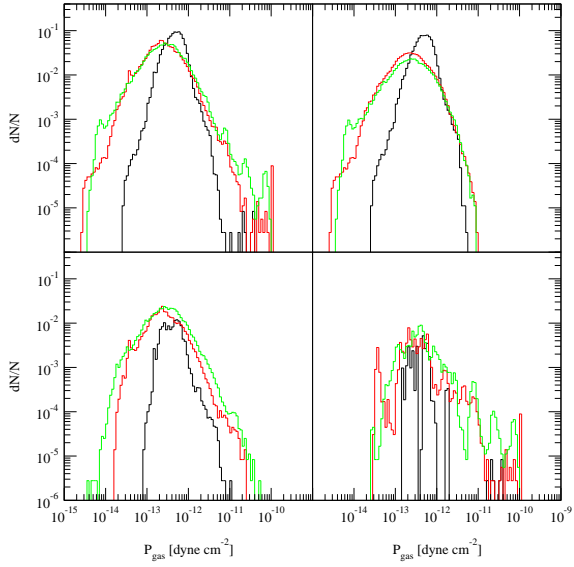,width=0.5\textwidth,angle=0} 
\caption{\label{fig:strat-pdf-rate} Volume-weighted PDFs of pressure
from the stratified models at different SN rates of one (black, model
S2), six (green, model S3), and ten (red, model S4) times the galactic
rate for the full distribution (upper left), and for cool gas with
$\log T < 3.9$ (upper right), warm gas with $3.9 < \log T < 5.5$
(lower left), and hot gas with $\log T > 5.5$ (lower right) in the
1.25~pc resolution case.  
}
\end{figure}

\begin{figure} 
\centerline{\hbox{
\psfig{file=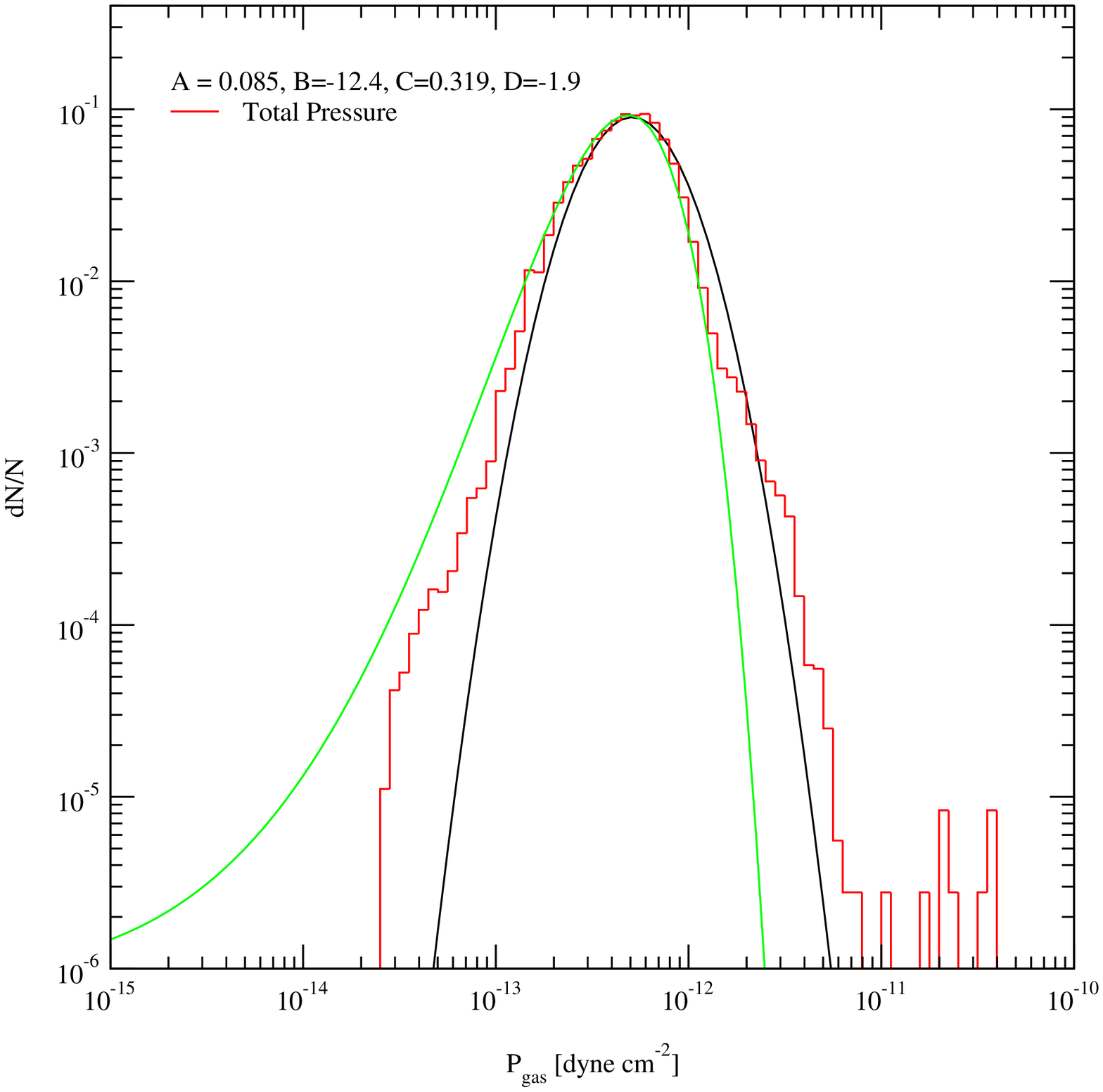,width=0.5\textwidth}
\psfig{file=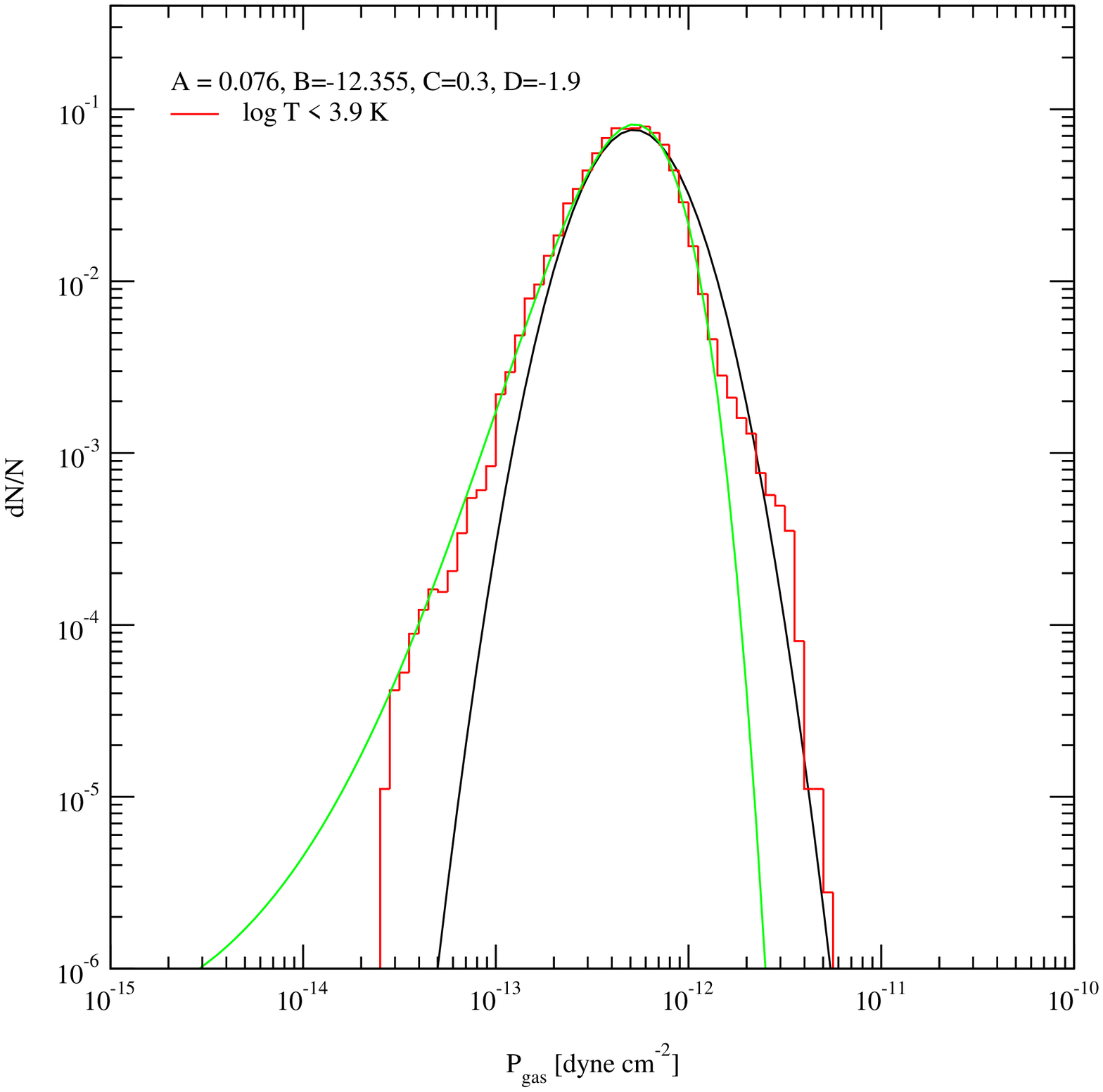,width=0.5\textwidth}
}}
\centerline{\hbox{
\psfig{file=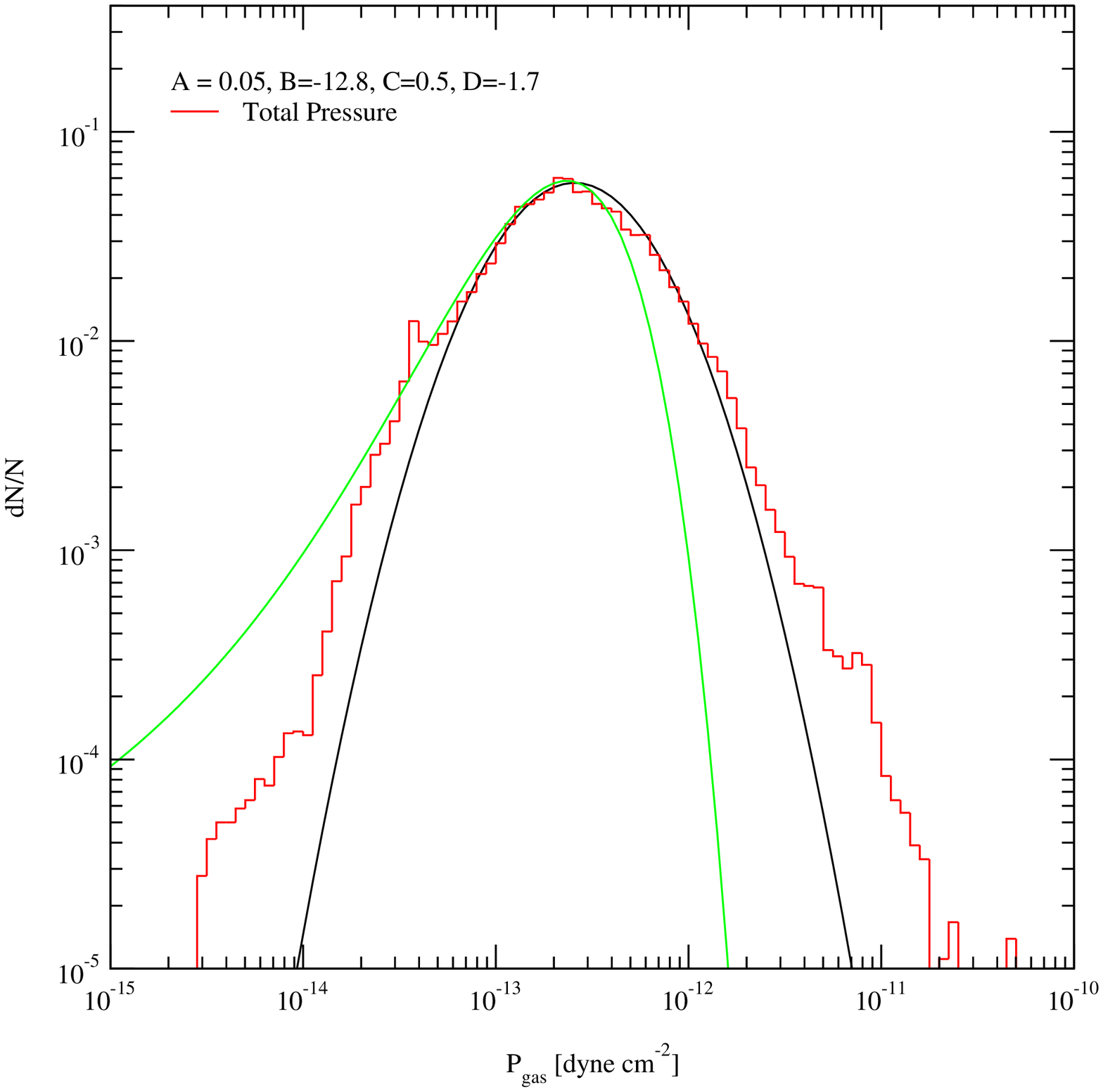,width=0.5\textwidth}
\psfig{file=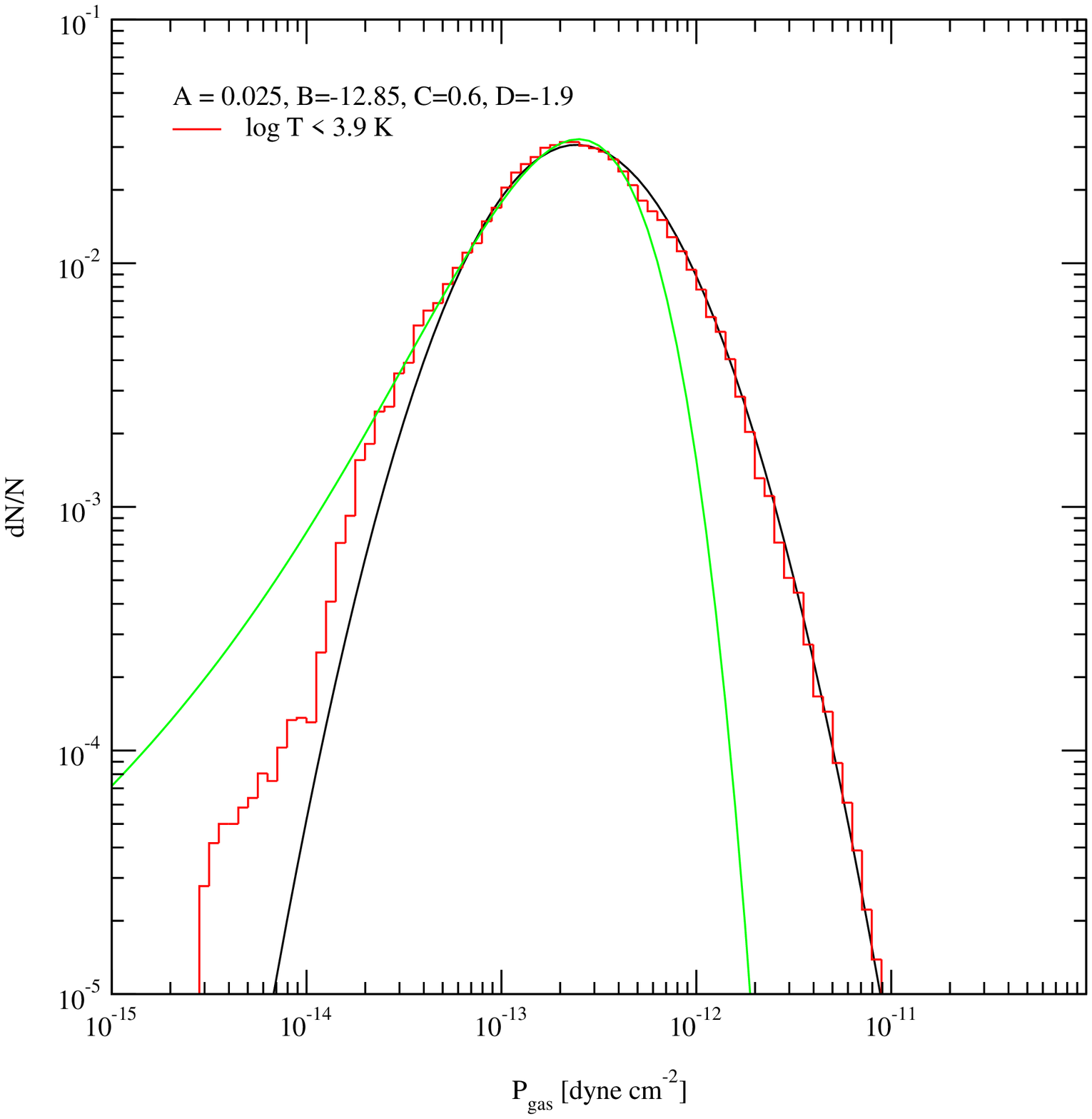,width=0.5\textwidth}
}}
\caption{\label{fig:strat-fits} Best fit log Gaussians (black) and
tilted log Gaussians following PV98 (green) for the total PDFs (left)
and PDFs for cool gas with $\log T < 3.9$ (right) for SN rates of (a)
Galactic (model S2) and (b) six times Galactic (model S3).  The best
fit parameters for the tilted Gaussian as described in
equation~(\ref{eq:fit}) are given in each case.
}
\end{figure}

We have run the stratified case with three different SN rates, as
shown in Figures~\ref{fig:strat-cut-rate}
and~\ref{fig:strat-pdf-rate}.  This allows us to quantitatively test
how well our results agree with the analytic theory of PV98 for purely
isothermal or polytropic turbulence.  We fit two different functions
to our pressure PDFs.  First, we tried a simple log-Gaussian
commensurate with the form of equation~(\ref{eq:pdfiso}) for an
isothermal gas,
\begin{equation}\label{eq:gfit}
{\cal P}(x) dx = A \exp [-(x-B)^2/C^2],
\end{equation}
where $x = \log_{10} P$ as before.  Then we fit a more complex
function capturing the predicted behavior of a polytropic gas with
polytropic index $\gamma$,
\begin{equation}\label{eq:fit}
{\cal P}(x)dx = A \exp \left[ -\frac{(x-B)^2}{C^2} \exp
\left\{\frac{(\gamma - 1) (x-B)}{\gamma\epsilon}\right\} - D(x-B)
\right].
\end{equation}
In both cases, the pressure dispersion is $\sigma_x = C/\sqrt{2}$.  We
find that the value of $\sigma_x$ never differs by more than 10\%
between the Gaussian and the tilted-Gaussian fits.  (This can be seen
by deriving the tilted Gaussian values from the values of $C$ given in
Figure~\ref{fig:strat-fits}.)  For simplicity, we therefore use only
the Gaussian values in what follows.

We would now like to compare the dispersion $\sigma_x$ that we find in
our Gaussian fits to the prediction of PV98 given in
equation~(\ref{eq:sigma}).  To do this we first need to derive an
effective central sound speed $c_0$ for our models. As an
approximation to $P_0$, we take the peak of our fits to the PDFs.
This is not quite right, due to the correction factor
$s_0=-0.5\sigma_s^2$, but for the relatively low values of $\sigma_x$,
and thus of $\sigma_s$, that we find in these models, it is probably
good enough.  We then find $\rho_0$ and $\gamma$ from the ridgeline at
low temperature of the distribution of points in the pressure-density
plane (Figs.~\ref{fig:strat-dscat} and~\ref{fig:mhd-dscat}), roughly
following the temperature where the cooling drops sharply in the
stratified hydrodynamical case, and the thermodynamic equilibrium
curve in the MHD case.  This procedure gives $c_0 \simeq
3$~km~s$^{-1}$ for the stratified models (corresponding to a
temperature of $10^3$~K and adiabatic behavior with effective $\gamma
= 5/3$), and $c_0 \simeq 6$~km~s$^{-1}$ for the MHD case (using the
effective $\gamma = 0.6$ from the slope of the curve).

We then use the rms velocities of each model, which we will discuss in
more detail in future work, to predict $\sigma_x$.  In
Table~\ref{tab:runs} we first compare the fits to the widths predicted by
the rms velocity for the full PDF, $\sigma_x(\mbox{total})$.  The fits
to the computational results are factors of five to ten narrower than
expected based on the rms velocity.  We then compared the computed
width of the PDF for only the cool gas with $\log_{10} T < 3.9$ to the
prediction based on the rms velocity of the cool gas,
$\sigma_x(\mbox{cool})$.  Here, the agreement for the Galactic SN rate is
excellent (although the numerical agreement is surely accidental), and
it only gradually degrades at higher driving rates.

The grounds for the strong disagreement between the fit and predicted
$\sigma_x(\mbox{total})$ seem pretty clear.  The high-velocity,
high-temperature gas is associated with young SNRs and thus has not
been fully incorporated into the turbulent flow.  The remaining
disagreement for the cold gas $\sigma_x(\mbox{cool})$ also has several
possible explanations: we are not using a polytropic equation of
state; our models are three-dimensional instead of one-dimensional;
and the distinction between the fully turbulent gas and the more
laminar young SNRs is not simply one of temperature.  Which of these
explanations is most important is less clear.  Nevertheless, we may
conclude that the rms Mach number of the cold gas does provide a
pretty reasonable first guess to the distribution of pressures in the
cold gas through equation~(\ref{eq:sigma}).

\section{Comparison to Observations}
\label{sec:observations}

\subsection{Ionized and Atomic Gas}

Evidence for a broad distribution of interstellar pressures had
already been noted as early as the work on excited levels of C{\sc i}
by Jenkins \& Shaya (1979) and Jenkins et al.\ (1983).  The point was
sharpened with the direct comparison of nearby pressures by Bowyer et
al.\ (1995), who compared the pressure of the interstellar medium
impinging on the heliosphere with the pressure of the interstellar
medium averaged over a 40~pc line of sight as measured with the {\em
Extreme Ultraviolet Explorer}.  Very recently, Jenkins \& Tripp (2001)
have used the STIS to extend the work on C{\sc i} with better resolved
data, concluding that the pressure varies by over an order of
magnitude both above and below the average value in a small fraction
of the gas.  Our models naturally explain these variations in the
context of a SN-driven interstellar medium.  We will endeavor to make
a more quantitative comparison soon.

Jenkins et al.\ (1983) compared their results to the analytic theory
of MO, described above in \S~\ref{sec:an-th}.  They found a
substantially greater column density of low-pressure material than
that predicted, and rather less high-pressure material.  As the
analytic theory could not predict pressures much less than average,
while our models show a broad range of rarefaction waves producing a
log-normal pressure distribution around the mean, the low-pressure
results appear consistent.  Jenkins et al.\ were observing emission
from excited states of C{\sc i} in low-temperature gas.  We find that
much or most of the high-pressure gas resides in the hot medium, so,
although we do predict a greater total volume of high-pressure gas
than was predicted by MO, we expect that most of it would have been
ionized, and therefore not observable by Jenkins et al.\ (1983),
explaining their low column densities of high-pressure material.

Bowyer et al.\ (1995) compared the pressure they derived from extreme
ultraviolet emission along a line of sight to an IRAS detected H~{\sc
i} cloud that was determined to be 40~pc away using photometry of
stars superposed on the cloud boundaries.  They found a pressure of
$P/k = $19,000~K~cm$^{-3}$ for the hot medium, with an absolute lower
limit of $P/k = $7,000~K~cm$^{-3}$.  They compared this to the value
for the pressure in the Local Cloud of 700--760~K~cm$^{-3}$ derived by
Frisch (1994) from scattering of solar He~{\sc i}~584\AA radiation
from helium flowing in from the cloud through the heliosphere.  McKee
(1996) argues that a more careful treatment of the unknown ionization
fraction could lead to a local pressure a factor of three higher,
reducing, but not eliminating, the discrepancy.  We find greater than
order of magnitude variations in our models, with pressures reaching
values as low as a few hundred K~cm$^{-3}$ in isolated regions in both
sets of models.  Interestingly, the lowest pressures are reached at
moderately low densities of order 0.1, while the density derived for
the Local Cloud by, for example, Quemerais et al.\ (1994) is
0.14~cm$^{-3}$.  

An example of a region in the hydrodynamical simulations with high and
low pressure regions intertwined in the manner suggested by Bowyer's
observations is the superbubble seen in Figure~\ref{fig:strat-cut} in the
region $200 < X < 500$ and $100 < Y < 300$.  In this region, even
local pressure equilibrium is lacking on scales of tens of parsecs, in
contrast to all multi-phase models.  Similar regions form regularly
over time.  In the MHD simulations, an example of a magnetically
supported cold cloud in a hot bubble can be seen in the bottom right
corner, showing that the explanation advanced by McKee (1996) is also
viable, though not unique.

Jenkins \& Tripp (2001) found that their results implied an effective
polytropic index in the cold gas of $\gamma > 0.9$, somewhat higher
than the $\gamma = 0.72$ derived by Wolfire et al.\ (1995) for this
gas.  The suggestion advanced to explain this is that the regions
being compressed may be smaller than the cooling length scale, and so
may begin to behave adiabatically.  An alternative explanation may be
drawn from the broad range of pressures at which gas cools in our
model: the (relatively sudden) pressurization may happen prior to
cooling, rather than to already cooled gas, as required by the
derivation of $\gamma$ by Jenkins \& Tripp.

The high pressures produced by turbulence in our models also may help
elucidate the origin of the opacity fluctuations seen in both H~{\sc
i} (Dieter, Welch, \& Romney 1976; Diamond et al.\ 1989; Faison et
al.\ 1998) and metal lines (Meyer \& Blades 1996; Watson \& Meyer
1996; Lauroesch et al.\ 1998, Meyer \& Lauroesch 1999).  These
fluctutions were described by Heiles (1997) as tiny-scale atomic
structure at overpressures of one to two orders of magnitude to
average pressure of the cold gas, which he took to be roughly $P/k =
4400$~K~cm$^{-3}$.  Deshpande (2000) suggested that they were due to
the spectrum of rather smaller density fluctuations at all scales
observed to be produced by the interstellar turbulence.  The high
pressures we observe in a small fraction of the interstellar gas, in
combination with the turubulent structure seen, appear qualitatively
consistent with these explanations.  Quantitatively distinguishing
between them will require a more careful treatment of the cold gas
physics and radiative transfer, which we defer to future work.  We do
note that computations of supersonic isothermal turbulence yield
structure functions qualitatively consistent with those required by
Deshpande (Mac Low \& Ossenkopf 2000; Ossenkopf \& Mac Low 2001).

\subsection{Molecular Gas}

Molecular clouds are observed to have broad linewidths suggesting that
they are subject to pressures as high as $10^5$~K~cm$^{-3}$.  Larson
(1981) was one of the first to suggest that the effective pressure was
due to the self-gravity of the cloud.  Since then, several authors
have found that most of the individual clumps in molecular clouds are
not in hydrostatic balance between turbulent pressure and
self-gravity, but rather are confined by an external pressure (Carr
1987; Loren 1989; Bertoldi \& McKee 1992).  The explanation offered
for this by Bertoldi \& McKee (1992) was that the entire cloud was
still subject to self-gravity, even though individual clumps were not,
and so the effects of self-gravity on large scales produced pressures
that confined the clumps.

In our models, we find pressures in high-density regions of order
$10^5$~K~cm$^{-3}$ in the absence of self-gravity, as shown in
Figures~\ref{fig:strat-dscat} and~\ref{fig:mhd-dscat}.  These
pressures are sufficient to confine observed clumps without invoking
self-gravity, suggesting that observed molecular clouds may be
primarily pressurized by the ram pressure of the turbulent flows in
which they are embedded rather than being self-gravitating objects.
Simulated observations of turbulent flows appear to suggest that
mass-linewidth relations thought to indicate that they are in virial
equilibrium may actually be due to the turbulence itself, or the
properties of the observations (Vazquez-Semadeni, Ballesteros-Paredes,
\& Rodriguez 1997; Ossenkopf, Ballesteros-Paredes, \& Heitsch 2001).

Ram pressure is a double-edged sword, however, that can destroy clouds
as easily as creating them.  This is consistent with the suggestion
that they are transient objects with lifetimes of under $10^7$~yr
first made by Larson (1981), and recently emphasized by
Ballesteros-Paredes, et al.\ (1999a) and Elmegreen (2000).  Relying on
ram pressures rather than self-gravity to confine observed clumps in
molecular clouds would also be consistent with the results of
simulations of hydrodynamical and MHD driven, self-gravitating,
isothermal turbulence that showed self-gravity only acting on small
scales, with turbulent flows dominating the large scales (Klessen et
al.\ 1999; Heitsch et al.\ 2001).  The same flows that confine and
destroy the clumps also drive the turbulence observed within them, as
the background flow stretches, twists, forms, and destroys dense
regions contained within it.

We note that the simple prescriptions we use for cooling below
$10^5$~K, where non-equilibrium ionization effects become important,
and more especially below $10^4$~K, where thermal instability is
physically expected to be present, are not realistic enough for us to
have any degree of confidence in the exact amounts of cold gas
produced at any given pressure.  However, our qualitative results,
produced with two rather different sets of physics, appear quite
robust, so we expect future work to refine the details rather than
substantially change our picture of molecular clouds forming in
transient, high-pressure, high-density regions produced by a
supersonic, turbulent flow.

\section{Summary}
\label{sec:summary}

We have examined the distribution of pressures predicted from two
different sets of three-dimensional simulations of a SN-driven ISM.
In one case we included the effects of vertical stratification, while
in the other we included a rather strong magnetic field and a
distributed heating function.  In all the simulations we examined, we
found a broad distribution of pressures over more than an order of
magnitude.  In the cool gas, with $\log T < 3.9$, the probability
distribution function of pressure takes a roughly log-normal form with
the width of the distribution proportional to the rms Mach number of
the cool gas.  This form can be derived from the work of PV98 on
density distributions, and markedly differs from the power-law
distribution predicted by previous work.  The higher the SN rate, the
broader is the variation of pressures from the mean in both high and
low directions.

The only isobaric regions are the interiors of young SNRs; these are
also the only laminar flow regions in an otherwise turbulent
environment.  Most of the mass is in the turbulent gas, however, as is
most of the volume even in models driven with ten times the Galactic
supernova rate (although questions of the relative strengths of
physical turbulent mixing and numerical diffusion will have to be
resolved before this conclusion is secure). 

We find a broad range of pressures, and a substantial fraction of
associated densities far from the thermal equilibrium values.  This
limits the predictive usefulness of phase diagrams based on thermal
equilibrium, although thermal equilibrium at the local pressure will
still be the mildly favored state.  Gas pressures appear to be
determined dynamically, and individual parcel of gas seeks local
thermal equilibrium at the pressure imposed on it by the turbulent
flow. as suggested by previous authors, the phase diagram is only
locally valid.  Isobaric thermal instability in such an environment
will lead to regions of the phase diagram being mildly disfavored, but
no more.  V\'azquez-Semadeni et al.\ (2000) and Gazol et al.\ (2001)
used two-dimensional simulations with more physically-motivated
cooling curves at lower temperatures to conclude that the effects of
the thermal instability will be barely visible in the overall
probability distribution functions.

Our results appear consistent with observations that have repeatedly
shown the ISM not to be isobaric, including those by Jenkins \& Shaya
(1979), Jenkins, Jura \& Loewenstein (1983), Bowyer et al.\ (1995),
and Jenkins \& Tripp (2001).  They suggest that, although heating and
cooling rates remain important for determining the local density and
temperature, they will not produce a global multi-phase medium because
of the wide range of pressures present and the dynamical processing of
the gas.  Rather, a more continuous distribution of densities and
pressures will always be present.

Inferences that molecular clouds must be gravitationally bound because
of their high observed confinement pressures are called into question
by these results.  Regions with densities approaching the overall
densities of GMCs, and pressures an order of magnitude above the
average interstellar pressure appear in our simulations even in the
absence of self-gravity.  This supports recent suggestions that
star-forming molecular clouds may be transient, turbulently-driven,
objects (Ballesteros-Paredes et al.\ 1999ab, Elmegreen 2000). (Objects
that do gravitationally collapse from large scales as described by Kim
\& Ostriker (2001) will also form molecular gas, but will quickly form
starburst knots in a burst of violent, unimpeded star formation.)

Even in the presence of a field rather stronger than observed for the
Milky Way, only small regions become magnetically supported, with
magnetic pressures substantially larger than thermal pressures.  The
exact filling factor of such regions will need to be determined by
more physically accurate studies in the future.  Low pressure regions
are much more frequently formed by turbulent rarefaction waves,
however, especially in gas with slightly lower than average density.

\acknowledgments We thank J. Ballesteros-Paredes, D. Cox,
E. B. Jenkins, C. F. McKee, T. Passot and E. V\'azquez-Semadeni for
valuable discussions, clarifications of their work cited here, and
useful comments on drafts.  M-MML and MAA acknowledge support from NSF
CAREER grant AST 99-85392, while DSB and JK acknowledge NSF grants
CISE 1-5-29014 and AST 00-98697.  Some computations presented here
were performed at the National Center for Supercomputing Applications,
which is also supported by the NSF.  This research has made use of
NASA's Astrophysics Data System Abstract Service.

\clearpage
\begin{center} {\Large Figures} \end{center}


\clearpage
\begin{deluxetable}{llrlllllll}
\tablecaption{Properties of the models. \label{tab:runs}}
\tablewidth{0pt}
\tablecolumns{10}
\tablehead{
\colhead{} & \colhead{} & \colhead{} & \colhead{$\Delta x_{\rm min}$}
& \colhead{$B_0$} & 
\multicolumn{2}{c}{Fit\tablenotemark{b}} & \colhead{} & 
\multicolumn{2}{c}{Predicted\tablenotemark{c}} \\
\cline{6-7} \cline{9-10} \\
	 \colhead{model}
        &\colhead{type}
	&\colhead{$\tau_{\rm SN}$\tablenotemark{a}}
	&\colhead{(pc)}
	&\colhead{($\mu$G)}
        &\colhead{$\sigma_x(\mbox{total})$}
        &\colhead{$\sigma_x(\mbox{cold})$}
        &\colhead{}
        &\colhead{$\sigma_x(\mbox{tot})$}
        &\colhead{$\sigma_x(\mbox{cold})$}                         
}
\startdata
S1 & strat & 1  & 2.5  & 0 & \nodata & \nodata & & \nodata &
\nodata \nl 
S2 & strat & 1  & 1.25 & 0   & 0.22    & 0.22    & & 0.99 & 0.22 \nl
S3 & strat & 6  & 1.25 & 0   & 0.35    & 0.39    & & 4.7  & 0.52 \nl
S4 & strat & 10 & 1.25 & 0   & 0.39    & 0.39    & & 5.7  & 0.61 \nl
M1 & mhd   & 12 & 3.13 & 5.8 & \nodata & \nodata & & \nodata &
\nodata \nl  
M2 & mhd   & 12 & 1.56 & 5.8 &  0.32   & \nodata & & 2.4  & \nodata \nl  
\enddata
\tablenotetext{a}{SN rate in terms of the Galactic SN rate}
\tablenotetext{b}{Dispersions of pressure derived from log-Gaussian fits to
numerical results}
\tablenotetext{c}{Dispersions predicted by equation~(\ref{eq:sigma})}

\end{deluxetable}

\begin{deluxetable}{lc}
\tablecaption{Cooling Times \label{tab:cool}}
\tablewidth{0pt}
\tablecolumns{2}
\tablehead{
\colhead{$\log_{10} T$} & \colhead{$t_{\rm cool} (n/1\mbox{
cm}^{-3})^{-2}$}  \\
\colhead{(K)} & \colhead{(yr)}
}
\startdata
3  &   1.7(5) \nl
4  &   6.0(3) \nl 
5  &   1.6(3) \nl
6  &   4.4(4) \nl
7  &   1.2(6) \nl
\enddata
\end{deluxetable}

\end{document}